\documentclass[3p,times]{elsarticle}

\usepackage{ecrc}
\usepackage{anyfontsize}

\volume{00}

\firstpage{1}

\journalname{Communications in Nonlinear Science and Numerical Simulation }

\runauth{Hasmi \& Susanto}

\jid{CNSNS}

\jnltitlelogo{CNSNS}

\CopyrightLine{2011}{Published by Elsevier Ltd.}

\usepackage{amssymb}

\usepackage[figuresright]{rotating}
\usepackage{amsmath}

\usepackage{amsmath}
\usepackage{amssymb}
\usepackage{amsbsy}
\usepackage{makeidx}
\usepackage{epsf}
\usepackage{graphicx}
\usepackage{amsfonts}

\usepackage{subfigure}
\usepackage{epstopdf}
\usepackage{mathrsfs}
\usepackage{xcolor}

\begin{document}


\dochead{}

\title{Model-free Forecasting of Rogue Waves using Reservoir Computing}

\author{Abrari Noor Hasmi}
\ead{100060615@ku.ac.ae}

\author{Hadi Susanto\corref{cor1}}
\ead{hadi.susanto@yandex.com}

\address{Department of Mathematics, Khalifa University of Science and Technology, PO Box 127788, Abu Dhabi, United Arab Emirates}
  \cortext[cor1]{Corresponding author}

\date{\today}
\begin{abstract}
Recent research has demonstrated Reservoir Computing's capability to model various chaotic dynamical systems, yet its application to Hamiltonian systems remains relatively unexplored. This paper investigates the effectiveness of Reservoir Computing in capturing rogue wave dynamics from the nonlinear Schr\"{o}dinger equation, a challenging Hamiltonian system with modulation instability. The model-free approach learns from breather simulations with five unstable modes. A properly tuned parallel Echo State Network can predict dynamics from two distinct testing datasets. The first set is a continuation of the training data, whereas the second set involves a higher-order breather. An investigation of the one-step prediction capability shows remarkable agreement between the testing data and the models. Furthermore, we show that the trained reservoir can predict the propagation of rogue waves over a relatively long prediction horizon, despite facing unseen dynamics. Finally, we introduce a method to significantly improve the Reservoir Computing prediction in autonomous mode, enhancing its long-term forecasting ability. These results advance the application of Reservoir Computing to spatio-temporal Hamiltonian systems and highlight the critical importance of phase space coverage in the design of training data.
\end{abstract}

\maketitle

\section{\label{chap4:introduction}Introduction}
Oceanic rogue waves have caused significant damage to ocean vessels and floating structures~\citep{Liu200757,didenkulova_catalogue_2020}. These waves are often described as walls of water that suddenly appear and disappear without a trace. The danger of rogue waves lies in their unpredictability, which makes it difficult for sailors to respond in time to mitigate their impact. They are defined as waves whose height is at least twice that of the surrounding waves (significant wave height).

There are two main views regarding the physics behind rogue waves~\cite{kharif_physical_2003}. The first suggests that rogue waves primarily arise from the linear superposition of waves, which, in rare instances, leads to strong constructive interference, resulting in exceptionally high waves~\citep{christou_field_2014,fedele_real_2016}. The second mechanism proposes a nonlinear process, where the interaction between the carrier wave and an unstable sideband wave induces modulation~\citep{onorato_freak_2001,onorato_modulational_2006,onorato_freak_2010}. A prototype of this nonlinear approach is the universal nonlinear Schr\"odinger equation (NLS) and its variants. NLS is a widely applicable equation in physics, modeling modulation instability and explaining rogue wave phenomena in a variety of fields, including ocean waves, optics, plasma, and atmospheric science, among others~\cite{Akhmediev2009,moslem2011SurfacePlasma,veldesElectromagneticRogueWaves2013,yangAnalysisRogueWaves2021,li2024CharacteristicsCertain,liu2025RogueWaves,slunyaevRogueWavesSea2023}.

The prediction of extreme waves, or rogue waves, has been a challenging task due to their nonlinearity and rare occurrence. The goal of rogue wave forecasting is to make short-term predictions before the event occurs. Prediction typically involves numerically solving wave envelope equations, such as the NLS or its variants, using numerical methods~\citep{onorato_freak_2001,klein_deterministic_2020}. Alternatively, researchers have sought to reduce the computational demands of numerical simulations by using reduced-order methods~\citep{Cousins2019,Farazmand2017} or machine learning techniques~\citep{kagemoto_forecasting_2020,kagemoto_forecasting_2022,breunung_data-driven_2023}.

Reservoir Computing (RC) is a specific variant of the recurrent neural
network (RNN) in machine learning~\citep{Jaeger2001}. RC can be perceived
as a three-layer RNN: an input layer, an internal layer state
(a reservoir), and an output layer. In contrast to the conventional RNN, the
reservoir parameters are fixed, sparse, and random. Surprisingly,
despite having a simplified learning procedure compared to RNNs,
RC has been reported to be more powerful~\citep{Jaeger2004,Lukosevicius2009a,Chattopadhyay2020}. The framework
has successfully modeled both short-term and long-term dynamics of chaotic
systems and captured its attractor \citep{pathak_using_2017,Pathak2018, Doan2021, Rohm2021a, gauthier_learning_2022}. An extension of RC for
learning spatiotemporal problems was presented in \citep{Pathak2018,Chattopadhyay2020,Huhn2021,Jiang2019,Pandey2020,yao2022LearningOcean,bonas2024CalibratedForecasts}.
Recent studies~\citep{pathak_hybrid_2018,Wikner2020} have introduced knowledge-based models within the RC framework. These advances highlight the versatility and potential of RC in addressing a wide range of spatiotemporal challenges. Another motivation for RC lies in its potential for physical implementation, as reviewed in~\cite{tanakaRecentAdvancesPhysical2019,liang2024PhysicalReservoir,everschor-sitte2024TopologicalMagnetic}, which has the potential to achieve both computational speedup and improved energy efficiency.

This research investigates the prediction and forecasting capabilities of RC to predict short-term nonlinear dynamics of integrable systems modeled by the NLS equation. Despite various studies on RC in chaotic 
spatiotemporal systems, little attention has been given to its application to Hamiltonian counterparts. An investigation into RC's ability to replicate the phase plane of Hamiltonian systems is reported
in~\cite{zhangLearningHamiltonianDynamics2021}, although this study is limited to low-dimensional systems. Only~\citep{Jiang2019} attempted to simulate NLS using RC, focusing on \textit{periodic} Akhmediev breathers,
Kutnezov--Ma solitons, and second-order periodic breather collisions. Despite this insight, the similarity between the training and testing data in~\citep{Jiang2019} raises questions about the model's generalizability.

We highlight key differences between dissipative and Hamiltonian chaos. In dissipative systems, phase space contracts, and the dynamics eventually settle onto a lower-dimensional structure known as an attractor. In contrast, the chaos observed in the focusing NLS equation arises from the presence of homoclinic orbits. The stable and unstable manifolds intersect transversally in phase space when these orbits are perturbed. However, due to the conservation of phase space volume in Hamiltonian systems, the trajectory evolves more complexly, passing near both the stable and unstable manifolds without settling, which leads to chaotic behavior. Unlike dissipative chaos, where the trajectory is confined to a lower-dimensional attractor, chaos in Hamiltonian systems typically spans a higher-dimensional region of phase space. This fundamental difference implies that dissipative system trajectories are concentrated around lower-dimensional attractors during machine learning model training. In contrast, in Hamiltonian chaos, the trajectory spans a more intricate and higher-dimensional phase space, making it harder for models to learn the full dynamics. The challenge is further compounded by the RC hidden layer, which is inherently dissipative and tends to contract trajectories toward attractors, making it especially difficult to sustain accurate, long-term predictions in Hamiltonian systems.

The primary objective of this research is to advance the understanding of RC's forecasting capabilities for \textit{non-periodic} waves, particularly rogue waves. A key focus lies in evaluating its performance using testing data from a different initial condition, aiming to thoroughly assess its robustness and applicability in varying scenarios. While several studies have demonstrated RC's ability to predict unseen dynamics in chaotic dissipative systems \citep{pershin_training_2023, Rohm2021a, Kim2021}, its performance in other systems, particularly Hamiltonian systems,remains largely unexplored due to the absence of an attractor. Our main contribution to the project is the revelation that RC can provide reasonable predictions, even when operating within different phase spaces in the integrable spatiotemporal system. Furthermore, we introduce a modified forecasting approach that incorporates real data assimilation, significantly improving RC’s predictive performance. This result paves the way for future research to better understand and improve the long-term predictive capabilities of RC within Hamiltonian systems.

This report is organized into the following sections. Section
\ref{chap4:Method} presents a brief introduction to parallel RC and details the training and testing procedures used in the study. In Section \ref{sec:SIMULATION}, we provide a detailed explanation of the training data and the prediction results. The main focus of our study is explained in Section \ref{sec:MI_Breather} and Section \ref{sec:ow_simulation}. In these sections, we employ different testing data to assess the generalizability of the RC model trained in the previous section for higher-order breathers and random ocean waves. Section \ref{sec:auto_obs} discusses procedures to extend the prediction horizon in autonomous prediction. 
We conclude in Section \ref{chap4:sec:CONCLUSION} and provide an overview of the future research direction. 

\section{Reservoir Computing Framework\label{chap4:Method}}
This section begins with a concise overview of RC. Following this introduction, we justify the specific RC architecture chosen for our investigation. Subsequently, we will detail our training and testing procedures in the following subsections.
 
\subsection{Parallel Reservoir Computing \label{chap4:subsec:Reservoir-Computing}}

The main principle of RC is to separate internal states (the reservoir) and readout learning. This distinguishes RC from other RNN architectures, which typically train both the states and the output parameters. In RC, only the readout layer is trained, making the learning process efficient and straightforward.

The following dynamical system describes the general form of RC: 
\begin{align}
x_{j} & =f(x_{j-1},u_{j-1}), \label{eq:Internal_State}\\
\hat{y}_{j}&=g(x_j).\label{eq:Output}
\end{align}
We will refer to the pair $(f,g)$ as an RC, where the input states are denoted as $u_{j} \in \mathbb{R}^{d_{I}}$ and the hidden states as $x_j \in \mathbb{R}^{d_{S}}$. As evident from Eqs.\ \eqref{eq:Internal_State}--\eqref{eq:Output}, RC is driven by a sequence of input states $\mathbf{u} = (\ldots, u_{-1}, u_{0})$, complemented by an initial hidden state $x_{-\infty}$. We denote $\mathbf{x} = (\ldots, x_{-1}, x_{0})$ and write $\mathbf{x} = f(\mathbf{u}, x_{-\infty})$ if each element of the sequence $\mathbf{x}$ is obtained by applying the operation in \eqref{eq:Internal_State} to the input sequence $\mathbf{u}$. Consequently, the output sequence is written as $\mathbf{y} = (f,g)(\mathbf{u})$. 

When the effect of the initial hidden state $x_{-\infty}$ becomes negligible after a certain number of iterations, implying that the hidden state depends solely on the input sequence, the dynamical system exhibits the Echo State Property (ESP)~\citep{Jaeger2004}. Technically, if $\mathbf{x} = f(\mathbf{u}, x_{-\infty})$ and $\tilde{\mathbf{x}}  = f(\mathbf{u}, \tilde{x}_{-\infty})$ are two sequences of hidden states corresponding to different initial conditions, the ESP condition implies that $x_0 = \tilde{x}_0$. 

In this study, we adopt a simple reservoir model, originally proposed by~\citep{Jaeger2001}:
\begin{align}
f(x_{j-1},u_{j-1})&=\tanh\left(W_x x_{j-1}+W_{u} u_{j-1}\right), \label{eq:Internal_F}\\
g(x_j) &= W_{o} x_{j}.\label{eq:Output_G}
\end{align}
Here, $\tanh$ acts as the activation function, applied element-wise. The matrix $W_x \in \mathbb{R}^{d_{S} \times d_{S}}$ serves as the adjacency matrix, which is sparsely connected, with an equal number of non-zero elements in each row and randomly valued to ensure the reservoir is rich enough to model nonlinear dynamics. The input coupling matrix $W_u \in \mathbb{R}^{d_{S} \times d_{I}}$ is a random and sparse matrix with entries uniformly distributed between $[-\alpha, \alpha]$. Notably, RC requires the dimension of the hidden state to be much larger than the input dimension, i.e.,~$d_S \gg d_I$. The output matrix $W_o \in \mathbb{R}^{d_O \times d_S}$ is learned by minimizing a loss function, which will be detailed in the next subsection.

In the absence of reservoir input, the ESP property in Eq.\ \eqref{eq:Internal_F} is ensured by rescaling the matrix so that the matrix norm of the adjacency matrix $\| W_x\| < 1$ or by constraining the largest spectral radius $\rho(| W_x| )$ to be less than 1 \citep{manjunath_echo_2013}. Here, $| W_x| $ represents the matrix obtained by taking the absolute values of the elements in $W_x$. However, this threshold should be viewed as a general guide rather than a strict constraint in an input-driven system. Previous studies, such as \citep{Jiang2019, Lukosevicius2009a}, have reported that even when this condition is violated in an input-driven system, the ESP property is still observed.

The choice of RC, as described by Eqs.\ \eqref{eq:Internal_F}--\eqref{eq:Output_G}, takes into account the symmetry inherent in the underlying data to be simulated. Specifically, the RC exhibits symmetry such that $(f,g)(\mathbf{u}) = -(f,g)(-\mathbf{u})$. The importance of preserving the symmetry of the RC output, which mirrors the symmetry of the original system, has been emphasized in previous works~\citep{Pathak2018,Barbosa2021}. Acknowledging and preserving this symmetry is crucial for ensuring the fidelity of the simulated dynamics and improving the ability of the RC model to accurately capture the underlying patterns and behaviors of the system being studied.

In this study, we implement the parallel RC approach introduced in~\citep{Pathak2018, Wikner2020}. This method divides the spatial domain into multiple RCs, each solving a subset of the problem independently. The rationale for dividing the data in this manner is that the exchange of information between distant spatial grids is not immediate; hence, creating RCs with connections between faraway physical points is unnecessary. However, to compensate for the lack of direct connections between distant grids, input points have overlapping regions between adjacent RCs. This strategy improves the efficiency of the RC model in capturing spatial dynamics. Additionally, the parallel RC model inherits spatial invariance from the underlying model, as described in the method introduced by~\citep{Barbosa2021}, which will be further elaborated in the following subsection.

\subsection{Training Procedure}

\begin{figure}[tb]
\includegraphics[width = \columnwidth]{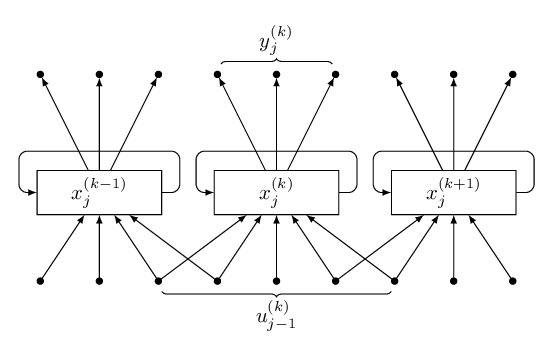}
\caption[Illustration of the parallel RCs architecture.]{Illustration of the parallel RCs architecture. The spatial grid is divided into overlapping grids $u^{(k)}$ as input to RCs and non-overlapping grids $y^{(k)}$ as the output. } \label{fig:parallel_RC}
\end{figure}

Before explaining the training procedure in more detail, we must clarify the notation used for the parallel RC in our approach. We will use a subscript $j$ attached to variables to denote time-dependent quantities, and we will use a superscript with parentheses $(k)$ to denote different parallel RCs. Let the dimensions of the spatial and temporal grids of the data be denoted by $L$ and $N$, respectively. Let $M$ represent the number of parallel RCs. The dimension of the output grid for each RC is then given by $m = L/M$. We generate two types of spatial grids: an overlapping grid $\mathbf{u}^{(k)} = \begin{pmatrix} u_{-M}^{(k)}, \ldots, u_{-1}^{(k)} \end{pmatrix}$, which serves as the reservoir input, and the partitioned grid, which is used as the RC output $\mathbf{y}^{(k)} = \begin{pmatrix} y_{-M+1}^{(k)}, \ldots, y_{-1}^{(k)}, y_{0}^{(k)} \end{pmatrix}$. The dimension of each term is given by $u_{i}^{(k)} \in \mathbb{R}^{m+2l}, y_{i}^{(k)} \in \mathbb{R}^{m}$, where $l$ represents the overlapping grids between RCs. This parallel RC setup is illustrated in Fig.\ \ref{fig:parallel_RC}.

The construction of RCs begins with the random generation of the matrices $W_{u}^{(k)}$ and $W_x^{(k)}$. Adjusting the spectral radius of $W_x^{(k)}$ is crucial during this process. The hidden states $\mathbf{x}^{(k)}$ are then computed based on the input data $\mathbf{u}^{(k)}$, as specified in Eqs.\ \eqref{eq:Internal_State} and \eqref{eq:Internal_F}. As noted in~\citep{Jaeger2001, Rohm2021a}, introducing noise into the input data enhances the autonomous prediction ability. Thus, the input states $\mathbf{u}^{(k)}$ are perturbed by Gaussian noise with variance $\sigma$. Note that the output states $\mathbf{y}^{(k)}$ are kept free of noise.

The loss function for each parallel reservoir is given by ridge (or $L_2$) regression: 
\begin{equation}
E^{(k)}(W_{o}^{(k)})=\sum_{j=-N+N_{w}}^{0}\left\Vert W_{o}^{(k)}x_{j}^{(k)}-\hat{y}_j^{(k)}\right\Vert_2 ^{2}+\beta\left\Vert W_{o}^{(k)}\right\Vert_F^{2},\label{eq:Cost_Function}
\end{equation}
where $N_{w}$ is the number of initialization steps used to "warm up" the RCs. The first
$N_{w}$ hidden states are discarded to eliminate any dependencies on initialization. The terms $\hat{y}_j^{(k)}$ represent the true output data, and $\beta$ is the Tikhonov parameter that controls the regularization strength of the least-square problem. The norm of the matrix $W_o$ is computed using the Frobenius norm. 

A special case of parallel RC arises when not only the adjacency matrix ($W_x^{(k)} = W_x$) and input matrix ($W_{u}^{(k)} = W_{u}$) are the same, but also the output matrix~\citep{Barbosa2021}. This procedure can be advantageous if the underlying system has translation invariance. In such cases, the output matrix should be independent of the particular partition of the training data. This situation often arises when the training data is generated from a numerical simulation of a partial differential equation using a uniform grid size, as in our problem. Implementing this procedure preserves the inherent symmetry in the system and improves computational efficiency, as it requires solving only one matrix equation for all parallel RCs. Technically, the loss function in Eq.\ \eqref{eq:Cost_Function} then becomes:
\begin{equation}
E({W_{o}})=\sum_{k=1}^{M}\sum_{j=-N+n_{w}}^{0}\left\Vert W_{o}x_{j}^{(k)}-\hat{y}_j^{(k)}\right\Vert_2 ^{2}+\beta\left\Vert W_{o}\right\Vert_F ^{2}.\label{eq:cost_tran_inv}
\end{equation}

Let us define $\mathbf{X}^{(k)}$ and $\mathbf{\hat{Y}}^{(k)}$ as
matrices whose columns are the vectors $x^{(k)}_{j}$ and $\hat{y}^{(k)}_{j}$,
respectively, for $-N+N_w\leq j\leq 0$. Furthermore, we construct the matrix $\mathbf{X}$ by stacking the matrices $\mathbf{X}^{(k)}$ column-wise, and similarly for $\hat{\mathbf{Y}}$. The loss function \eqref{eq:cost_tran_inv} is then expressed in matrix form as: 
\begin{equation}
E({W_{o}})=\left\Vert W_{o}\mathbf{X}-\mathbf{\hat{Y}}\right\Vert_F ^{2}+\beta\left\Vert W_{o}\right\Vert_F ^{2}.\label{eq:cost_tran_inv_matrix}
\end{equation}

Thus, the parallel RC training problem becomes an ordinary least squares problem for $W_o$. The solution is obtained by setting the first derivative to zero, which leads to the following equation:
\begin{equation}
\left(\mathbf{X}\mathbf{X}^{T}+\beta\mathbf{I}\right)W_{o}^{T}=\mathbf{X}\mathbf{Y}^{T},    \label{eq:train_matrix}
\end{equation}
 where $T$ denotes the transpose of the matrix, and $\mathbf{I}$ is the identity matrix of appropriate dimensions. 

It is important to note that the structure of the matrix least squares problem for the general parallel RC is similar to Eq.\ \eqref{eq:train_matrix}. However, in the general case, we solve for each $W_o^{(k)}$ independently, considering the corresponding $\mathbf{X}^{(k)}$ and $\hat{\mathbf{Y}}^{(k)}$, rather than stacking them as $\mathbf{X}$ and $\hat{\mathbf{Y}}$.

\subsection{Testing Process and Quality Metric \label{subsec:Testing-Process}}

Two primary prediction modes are used to assess the predictive capability of RC~\citep{Pyle2021}. The
first mode is "teacher forcing" or one-step prediction, where the reservoir
is fed with test input data $\mathbf{u}_{+}^{(k)}=\begin{pmatrix}u_{0}^{(k)} & u_{1}^{(k)} &\ldots \end{pmatrix}$. Note that the testing input data are "ground-truth" data derived from the same partial differential equation as the training data, but with different times or initial conditions. The hidden states evolve according to Eqs.\ \eqref{eq:Internal_State}  and \eqref{eq:Internal_F}. The one-step prediction is obtained by applying Eqs.\ \eqref{eq:Output}  and  \eqref{eq:Output_G} using the learned output matrix $W_{o}$. It is important to note that, in this mode, the RC prediction $\hat{y}^{(k)}_{j}$ depends on the reservoir states $x^{(k)}_{j-1}$  but is independent of the previous time step's prediction, $\hat{y}^{(k)}_{j-1}$. This mode is valuable for assessing the model's generalization ability to unseen inputs. 

The second prediction mode is the generative or autonomous mode. In this mode, the reservoir is initially fed with ground-truth input data $\mathbf{u}_{+}^{(k)}$ until a specific time index $j_0$. Subsequently, the RC prediction at time $j_0$, $\hat{y}_{j_0}^{(k)}$, is used as the input states $\hat{u}_{j_0}^{(k)}$ to update the hidden states $x_{j_0+1}^{(k)}$ and generate the prediction $\hat{y}_{j_0+1}^{(k)}$. This iterative process continues until the desired prediction steps are achieved. For our parallel RC, there is a subtle difference between the RC prediction $\hat{y}_{j}^{(k)}$ and the feedback input $\hat{u}_{j_0}^{(k)}$, due to the overlapping grid. Specifically, the input $\hat{u}_{j_0}^{(k)}$ should be assembled using $\hat{y}_{j_0}^{(k)}$ and its adjacent predictions $\hat{y}_{j_0}^{(k \pm 1)}$, as illustrated in Fig.\ \ref{fig:parallel_RC}. It is important to note that in the generative mode, the error will propagate during subsequent predictions, making this prediction more challenging than the teacher-forcing mode. Although the generative mode is commonly used to assess RC performance, we aim to understand both types of prediction modes.

The result of our RC prediction is presented on an assembled grid, where each parallel RC prediction $\hat{y}_j^{(k)}$ for all $k$ is combined into a single prediction $\hat{y}_j$. We define the normalized root mean square error (NRMSE) as: 
\begin{equation}
     e_j = \frac{\|\hat{y}_j - y_j\|}{\|y_j\|}, 
\end{equation}
where $y_j$ represents the ground-truth result. 

The evaluation of autonomous prediction at a certain time index $j$ depends on the starting time index of the autonomous prediction, $j_0$. We clarify this by defining the prediction length $j_l$ as $j_l = j - j_0$, and the NRMSE for autonomous prediction as: 
\begin{equation}
     e_{j_0,j_l} = \frac{\|\hat{y}_{j_0+j_l} - y_j\|}{\|y_j\|}. 
\end{equation}
For autonomous prediction, a meaningful metric is the prediction horizon, defined as $PH(j_0,\varepsilon)=\min_{j_l}e_{j_0,j_l}\geq\varepsilon$. The prediction horizon represents the earliest time index of an autonomous prediction, starting at index $j_{0}$, where $e_{j_0,j_l}$ exceeds the tolerance $\varepsilon$. In this study, we set $\varepsilon=0.4$ as the threshold for acceptable error. We will loosely refer to the prediction length or horizon in terms of discrete time indices or physical time units.  

\section{RC Prediction for Higher-Order Rogue Wave \label{sec:SIMULATION}}
In this section, we present the results of applying the parallel RC model to predict the occurrence of rogue waves. We begin by explaining how we generate the training data. Then, we outline the implementation of our training procedure and discuss the results for both one-step and generative prediction. 

\subsection{Generation of Ground Truth Data\label{subsec:Simulation-Setup}}

We train a set of reservoirs to learn a simulation governed by the following focusing NLS equation in its dimensionless form: 
\begin{equation}
    i\frac{\partial\psi}{\partial t} + \frac{1}{2}\frac{\partial^{2}\psi}{\partial\xi^{2}} + |\psi|^{2}\psi = 0, \label{eq:NLS}
\end{equation}
where the amplitude envelope $\psi(\xi, t)$ is a function of propagation
distance $\xi$ and co-moving time $t$. 

A well-known analytic solution of the NLS equation is the Akhmediev breather~\citep{Akhmediev1987}, given by:
\begin{equation*}
    \psi(\xi,t)=\left[1+\frac{2(1-2a)\cosh\left(\gamma t\right)+i\gamma\sinh\left(\gamma t\right)}{\sqrt{2a}\cos\left(\Omega\xi\right)-\cosh\left(\gamma t\right)}\right]e^{it},
\end{equation*}
where the behavior of the solution is determined by the modulation parameter $a \in (0, 0.5)$. Two other parameters associated with $a$ are the fundamental wavenumber $\Omega=2\sqrt{1-2a}$ and the growth factor $\gamma=\sqrt{8a(1-2a)}$. 

Numerical simulations of the NLS rogue wave are prone to homoclinic chaos, a phenomenon that arises purely due to numerical factors \citep{herbst_numerically_1989}. Different numerical algorithms, orders, and truncation errors can produce variations in the occurrence of rogue waves~\citep{belic_different_2022}. For our simulation, we use a clean numerical simulation approach from~\citep{hu_risks_2020}, as detailed in Appendix~\ref{APP:Num_Method},  implemented with quadruple precision accuracy. The increase in precision significantly improves the reliability of the NLS simulation by delaying the onset of rogue waves induced by round-off errors. 

The initial condition for our simulation is a plane wave disturbed by a single harmonic mode, given by:
\begin{equation*}
\psi(\xi,t_0)  = A_{0}+2A_{1}\cos(\Omega \xi),
\end{equation*}
where $\Omega$ is the fundamental frequency corresponding to $a \approx 0.4802$ or $\Omega \approx 0.39799$.  With this choice, there are five unstable modes corresponding to frequencies $\Omega, 2\Omega, \dots,5\Omega$, with respective growth factors $(\gamma_1, \gamma_2, \gamma_3, \gamma_4, \gamma_5) = (0.39, 0.73, 0.958, 0.964, 0.199) $.  We set $A_{1}=10^{-4}$, while $A_{0}=\sqrt{1-2A_{1}^{2}}$. The simulation uses a fixed timestep $\Delta t=2\cdot 10^{-4}$. 

The length of the spatial domain is one spatial period, $[-L/2,L/2]$, with the primary wavelength $L=2\pi/\Omega$. The spatial domain consists of 256 spatial nodes, with periodic boundary conditions on the lateral sides. After the first formation of the Akhmediev breather, higher-order recurrences are created due to higher-order dynamics modes located in the modulation instability range~\citep{Yuen1978,Erkintalo2011a,Chin2015}. Thus, a complex recurrence pattern grows and collapses as modes with different growth factors and phases interact nonlinearly.

\subsection{Training and One Step Prediction\label{Subsec:Train}}
\begin{figure}
\begin{centering}
\includegraphics[width=0.9\columnwidth]{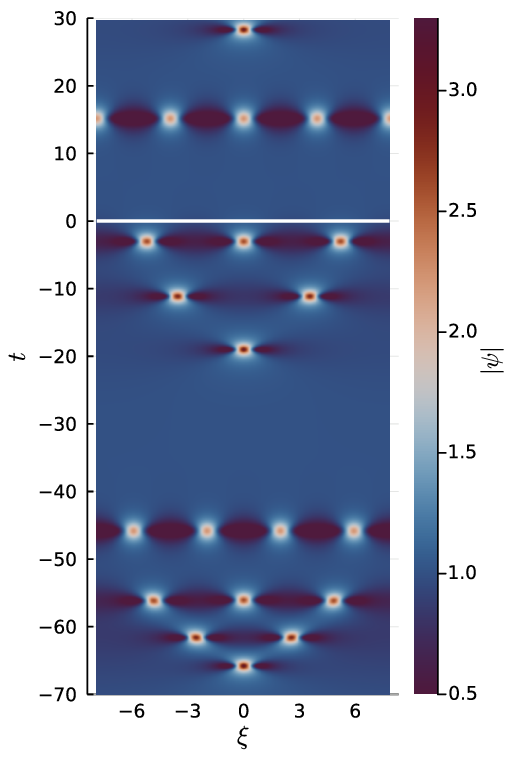}
\par\end{centering}
\caption[The NLS recurrence as training and testing data.]{The NLS recurrence as training and testing data. The data are generated
by disturbing plane waves with frequency $\Omega=0.39799$. The white line separates training and testing data.}\label{fig:train_NLS}
\end{figure}
This subsection discusses the implementation of our training procedure and its one-step prediction results for higher-order rogue wave data with recurrence. 

For our RC experiment, we separate the real and imaginary parts of the NLS, yielding 512 spatial nodes. We sample the data such that $\Delta t_{RC} = 5 \times 10^{-3}$ time units, equivalent to taking one sample every 25 time steps. Initially represented in quadruple precision, the simulated data is truncated to double precision. To eliminate the initial build-up of the first breather, we exclude 15 time units from the initial simulation. Subsequently, the next 70 time units of the data are used for training data, while the following 35 time units serve as testing data. The division between training and testing data is shown in Fig.\ \ref{fig:train_NLS}, with the white line indicating the boundary. We refer to this boundary as $t = 0$ in the following results.

\begin{figure}
\includegraphics[width=\columnwidth]{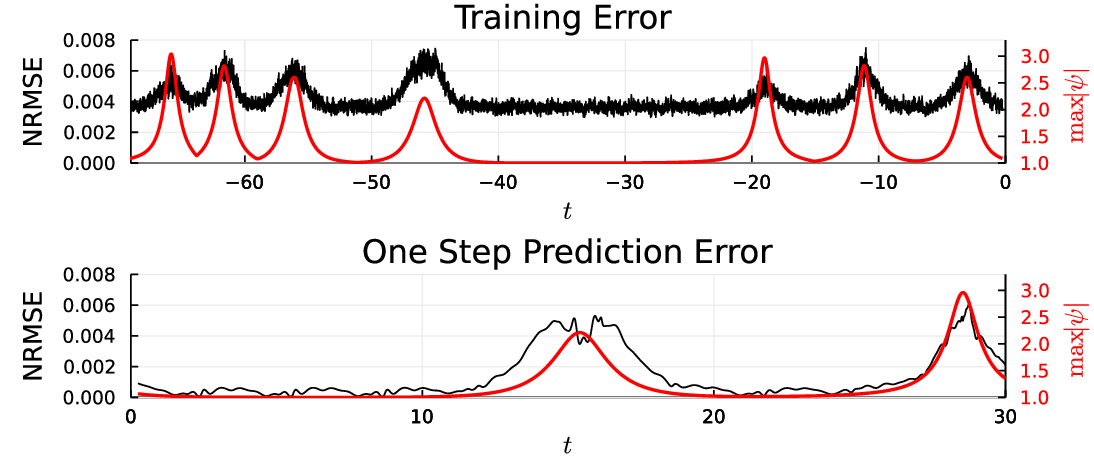}
\caption[Trained RC has low NRMSE for one-step forecasting.]{Trained RC has low NRMSE for one-step forecasting. The top and bottom panels show the time trace of training NRMSE and NRMSE of one-step testing prediction respectively. The black line represents the NRMSE, while the red line is the maximum wave envelope}
\label{fig:train_test_err}
\end{figure}

Parallel RC is constructed with $M = 64$  reservoirs, and each reservoir has $d_S = 800$ hidden nodes, resulting in $m = 8$. We incorporate two physical spatial nodes as overlapping areas, i.e.,~$l=4$, because the NLS is divided into real and imaginary parts. This results in an input dimension of $d_I = 16$. The hyperparameters were chosen by hand after observing that further increases in computational effort did not significantly affect the simulation or forecasting results.

For the training process, we set the regularization strength $\beta = 10^{-4}$ and use $100$ washout steps ($N_{w} = 100$). Remarkably, these hyperparameters are not highly sensitive. We introduce white Gaussian noise with a variance of $\sigma = 0.02$ only to the system input data $\mathbf{u}^{(k)}$. We observed an improvement in the autonomous prediction when noise was introduced, although the optimal noise value remains unclear~\citep{Jaeger2001}. Consistent with the finding in~\citep{Barbosa2021}, we observe that the translation invariance procedure significantly improved the autonomous prediction in our case. We will refer to this RC model as RC\_AB.

After training, we generate a teacher-forced reservoir by feeding the test data, $u_{j}^{(k)}$, for all $k$ and $j \geq 0$. Based on these reservoir states, we compute a one-step prediction at each time step $\hat{y}_{j+1}$ and compare it with the real test data $y_{j+1}$. In this mode, the one-step error at a given time instance does not propagate to the next prediction. The NRMSE for both training and one-step prediction are shown in Fig.\ \ref{fig:train_test_err}, along with the maximum wave envelope for comparison with Fig.\ \ref{fig:train_NLS}. The figure shows that the overall error in the training procedure is higher than in the one-step prediction, which can be attributed to high noise in the training data. The testing data also show that the error overshoots during the modulation instability, highlighting the difficulty of predicting nonlinear instability. Nevertheless, the error remains small, and an accurate one-step prediction is achieved even during the instability. 

\subsection{Autonomous Prediction\label{subsec:auto_pred}}

\begin{figure}
\includegraphics[width=\columnwidth]{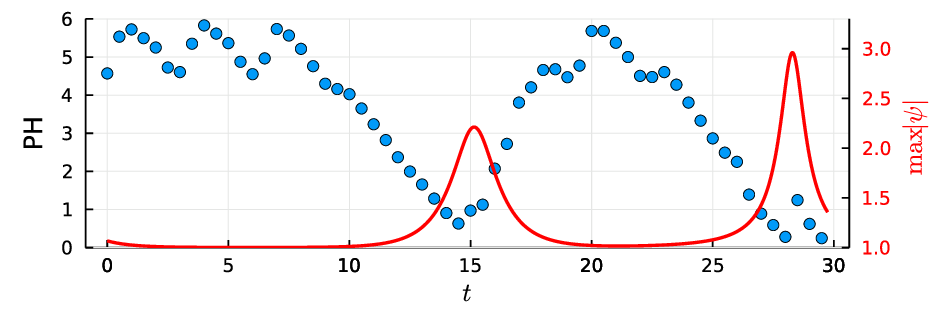}
\caption[The prediction horizon (PH) of autonomous prediction varies at different start times $t_0$.]{The prediction horizon (PH) of autonomous prediction varies at different start times $t_0$.  Close to the peak of modulation instability, PH becomes short.}
\label{fig:auto_result}
\end{figure}

We investigate the autonomous prediction performance of the RC on the test data. In this analysis, we vary the initial time of our autonomous prediction, $t_0$, and observe the prediction horizon. As noted previously, in each autonomous prediction, we always feed the RC with ground truth data before $t_0$ (see Sec.\ \ref{subsec:Testing-Process}). The results show that the reservoir has a long prediction horizon, as illustrated in Fig.\ \ref{fig:auto_result}. However, the RC has difficulty forecasting near the peak of the modulation instability. Nevertheless, the RC can detect rogue waves approximately two time units before their occurrence. 

Several RC studies use the Lyapunov time, defined as the inverse of the largest Lyapunov exponent, to scale the prediction horizon. However, these studies typically focus on chaotic dissipative systems with well-defined attractors, which correspond to systems with a well-defined Lyapunov exponent. In contrast, the NLS equation is an integrable Hamiltonian system, where the dynamics preserve phase space volume, thus precluding the existence of a low-dimensional attractor.

 As an alternative, we consider the maximum modulation instability growth factor, $\gamma_{\text{max}} = \gamma_4$, as the characteristic timescale. This choice is motivated by two main reasons: first, this maximum growth rate is equivalent to the largest Lyapunov exponent in the case of a plane wave solution, and second, the maximum exponential growth of a rogue wave follows this rate, analogous to the exponential divergence of trajectories in chaotic systems.  We support this claim by calculating the local largest Lyapunov exponent, as detailed in Appendix \ref{APP:LLE} . Given that $\gamma_{\text{max}} = 0.964$, the timescale shown in Fig.~\ref{fig:auto_result} is nearly aligned with this characteristic timescale.


\begin{figure}
\includegraphics[width=0.9\columnwidth]{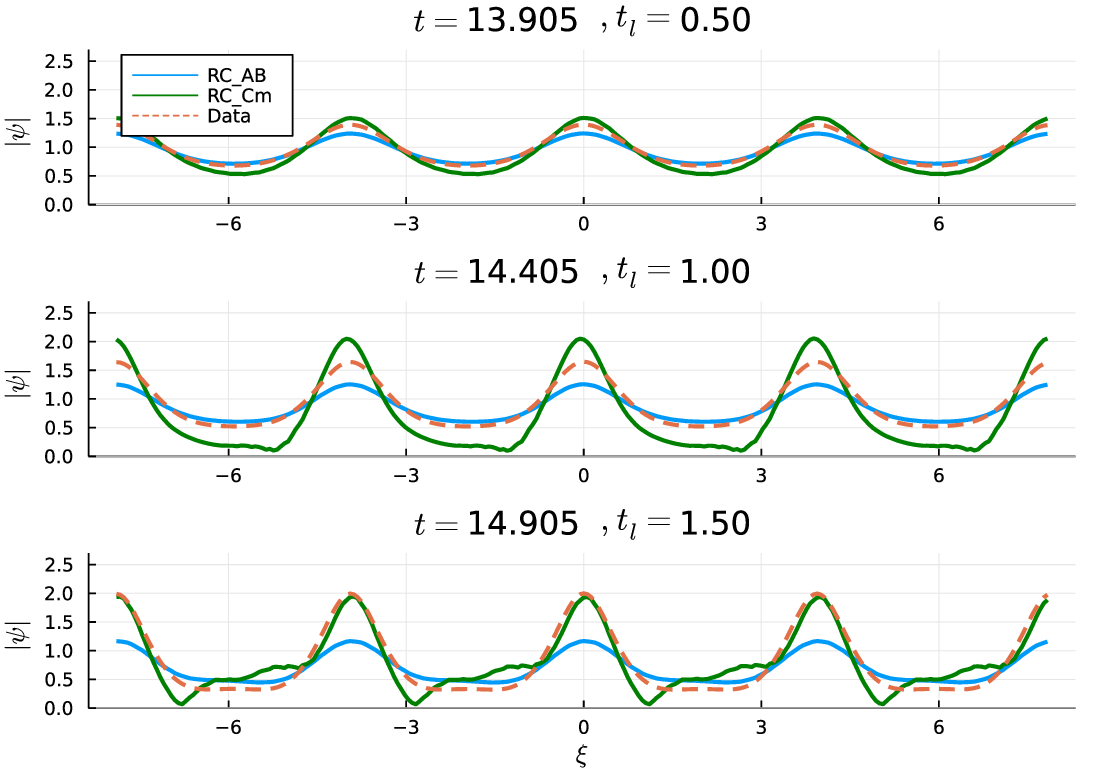}
\caption[Autonomous RC\_AB simulation for offset time instances.]{Autonomous RC\_AB simulation for offset time instances. Actual data are plotted for comparison. Note how the amplitude difference diverges. Nevertheless, RC\_AB can forecast the dynamics. The autonomous prediction starts at $t_{0}=13.4$. The green curves are from RC\_Cm, see Sec.\ \ref{sec:ow_simulation}.}
\label{fig:auto_profile}
\end{figure}

\begin{figure}
\includegraphics[width=0.9\columnwidth]{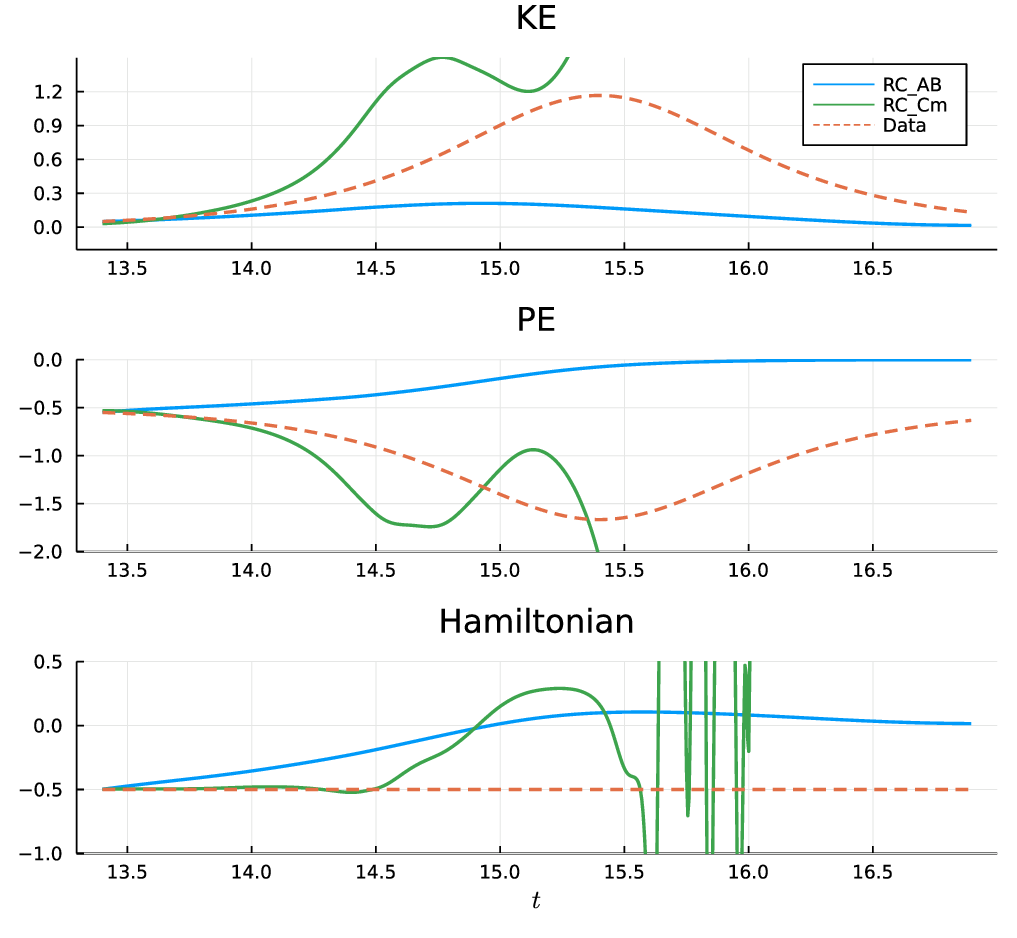}
\caption[Kinetic Energy (KE), Potential Energy (PE), and Hamiltonian (Ham) of the autonomous RC\_AB, RC\_Cm (see Sec.\ \ref{sec:ow_simulation}), and the NLS simulation.]{Kinetic Energy (KE), Potential Energy (PE), and Hamiltonian (Ham) of the autonomous RC\_AB, RC\_Cm (see Sec.\ \ref{sec:ow_simulation}), and the NLS simulation. The learned RCs behave like a dissipative system. The peak of the kinetic energy (around $t=15.4$) marks the breather occurrence. }
\label{fig:Hamiltonian}
\end{figure}

We further investigate the behavior of RC forecasting close to the peak of modulation instability. Starting the forecast at $t_0 = 13.4$, two time units before a breather peak, we run the autonomous prediction for two time units (400 timesteps). A comparison of the actual profile and the forecast result is shown in Fig.\ \ref{fig:auto_profile}. Although the overall profile can be captured by RC, the RC forecast underestimates the amplitude of the breather. To understand this, we compute the Hamiltonian as shown in Fig.\ \ref{fig:Hamiltonian}. The analysis indicates that the RC prediction gradually loses energy, with the peak kinetic energy occurring too early in the forecast. Since the modulation instability is driven by nonlinearity and energy conservation~\citep{Chin2015}, the dissipation of energy might explain the inaccuracy of the RC in propagating the model based on the NLS.

As a forecasting model, we do not expect RC to perfectly match the results of numerical simulations. Each step in RC introduces a deviation from the actual solution. In autonomous prediction, the previous predictions are used as input states, leading to error propagation over subsequent steps, which ultimately causes the model to fail in long-term forecasting. This is reflected in the deviations observed in Figs.\ \ref{fig:auto_profile} and \ref{fig:Hamiltonian}. Nevertheless, the critical question is how short or long the forecasting time is reliable, such that the model is useful. The answer to this depends on the specific application. However, since the NLS in Eq.~\eqref{eq:NLS} is in normalized time units, the corresponding prediction horizon in physical time can be useful for prediction, as in the real ocean wave simulation discussed in Section~\ref{sec:ow_simulation}. The prediction horizon in this section is considered "long" enough when compared with the prediction horizon for real ocean wave predictions. Moreover, we also consider another RC model, RC\_Cm (see Section~\ref{sec:ow_simulation}), which yields a significant improvement over the results of RC\_AB, as shown by the green curves in Figs.~\ref{fig:auto_profile} and \ref{fig:Hamiltonian}.

\section{Maximal Intensity Breather\label{sec:MI_Breather}}

Studies have shown that RC can handle unseen dynamics. For instance, Rohm et al.~\citep{Rohm2021a} reconstructed an unseen attractor, while Roy et al.\ \citep{royModelfreePredictionMultistability2022a} captured unknown attractors for different parameter values, and Pershin et al.~\citep{pershin_training_2023} explored prediction in the laminar-to-turbulent transition. In this work, we focus on simulating "unseen" dynamics of the NLS equation, specifically investigating the capability of a trained RC to simulate the occurrence of a maximum-intensity breather.

\begin{figure}[!tb]
\includegraphics[width=0.85\columnwidth]{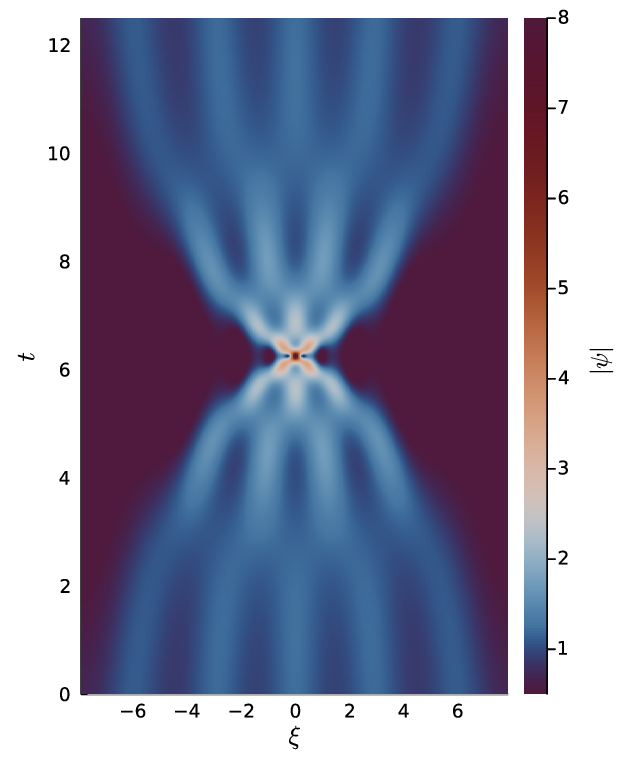}
\caption{A fifth-order maximum intensity breather as testing data.}
\label{fig:testing_data_AB}
\end{figure}

The concept of a maximum intensity breather, defined in~\citep{Chin2016a}, involves constructing an NLS solution that allows all unstable higher-order modes for a given wavenumber $\Omega$ to synchronize and reach maximum amplitude in a specific space and time. The initial condition is constructed using the Darboux transformation, as detailed in~\citep{akhmediev_extremely_1991, Kedziora2011}. In our simulation, we use $\Omega = 0.39799$, the same value as in Section \ref{sec:SIMULATION}, which is capable of generating a 5th-order maximum intensity breather. The resulting dynamics, depicted in Fig.~\ref{fig:testing_data_AB}, display five lobes that gradually converge into a single lobe with the maximum peak height $\left|\psi\right|_{\text{max}} = 1 + 2 \sum_{j=1}^{m} \sqrt{2a_j}$, where $a_j$ is the modulation parameter corresponding to unstable modes. This structure then collapses symmetrically over time. The maximal intensity breather data are generated using the same time and spatial grid as the NLS recurrence simulation in Section~\ref{subsec:Simulation-Setup}.

\begin{figure}[!tb]
\includegraphics[width=0.9\columnwidth]{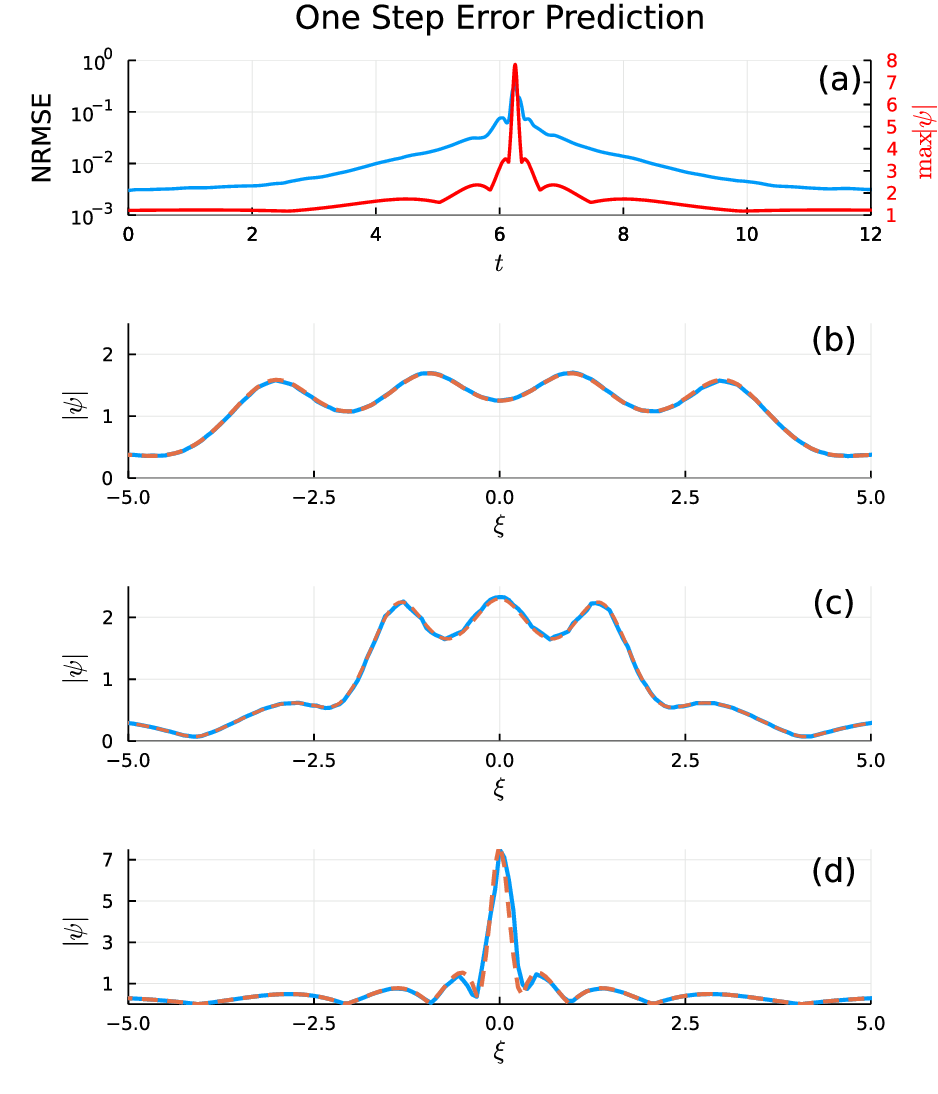}
\caption{RC\_AB can accurately forecast the dynamics of maximum intensity breather in Fig.~\ref{fig:testing_data_AB}
using one-step prediction, even though the RC was trained using data in Fig.~\ref{fig:train_NLS}. (a) The NRMSE of RC prediction is shown on a logarithmic scale. (b)-(d) Snapshots comparing RC prediction and maximal intensity breather data at $t=4.75, 5.75,6.25$.}
\label{fig:AB_test}
\end{figure}

We use the same reservoir, i.e., RC\_AB, as Sec.~\ref{sec:SIMULATION} without retraining. This means that we use the generated $W_{x},W_{u}$
along with the learned $W_{o}$  to generate hidden states $\mathbf{x}^{(k)}$  from the input testing data shown in Fig.~\ref{fig:testing_data_AB}.  We discard $N_w$ transient points before calculating one-step predictions, as presented in Fig.~\ref{fig:AB_test}. The NRMSE of one-step prediction is displayed in panel (a), with the error axis shown on a logarithmic scale. The figure shows that the one-step error is generally insignificant, except near the peak of the breather. Even though the multiple lobes are not present in the RC training data, the RC's one-step prediction profile is remarkably similar to the expected result, as observed in panels (b) and (c). Furthermore, although the RC prediction slightly underestimates the highest peak, as shown in panel (d), the RC successfully captures the overall shape of the highest peak breather.

\begin{figure}[tb]
\includegraphics[width=1\columnwidth]{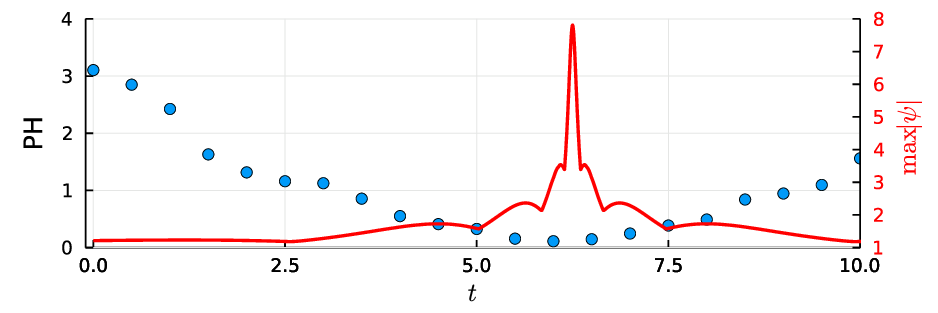}
\caption[RC can do a short time forecasting of propagation of maximal intensity breather. ]{RC can do a short time forecasting of propagation of maximal intensity breather. The blue dots represent prediction horizons. The red line represents the maximum of the wave envelope.}
\label{fig:auto_400}
\end{figure}

Figure \ref{fig:auto_400} shows the prediction horizon of the maximal intensity breather testing data, with a legend similar to Fig.~\ref{fig:auto_result}. Overall, the prediction horizon in the maximum intensity breather is shorter than in Section \ref{subsec:Testing-Process}. Despite the shorter prediction horizon, our RC can forecast the occurrence of extreme rogue waves with a lead time of 0.5 time units. 

The maximal intensity breather represents a path within the homoclinic orbit. It is important to note that the RC model can provide reasonable one-step predictions, even though the original training data do not directly traverse this trajectory. However, due to the inherent sensitivity of the homoclinic orbit, where any infinitesimal deviation leads to exponential divergence, RC struggles with long-term autonomous prediction. The rapid divergence of trajectories near the homoclinic orbit highlights the challenge of maintaining accurate forecasts beyond short horizons. 

\section{Random Ocean Wave Simulation\label{sec:ow_simulation}}
This section demonstrates the forecasting performance of RC in predicting the occurrence of rogue waves in random ocean waves. To that end, we synthesize long-crested irregular waves in deep water. The initial condition for the simulation follows the JONSWAP spectrum, with the detailed procedure given in Appendix \ref{App:JONSWAP}. The physical conditions correspond to deep water waves with a peak period $T_p = 8$ s and a significant wave height $H_s = 8$ m. According to the linear ocean wave dispersion relation, this corresponds to a water depth of at least 50 mand a significant wave steepness of $H_s / L_p = 0.08$, where $L_p$ is the wavelength at the peak period. We then generate several sets of data with different random phase realizations.

In addition to the RC model developed in Subsection \ref{Subsec:Train} (RC\_AB), we also consider another model referred to as RC\_Cm. The RC\_Cm model utilizes the same training data as RC\_AB along with one of the simulations of random waves. Other parameters remain fixed, including the matrices $W_x$ and $W_u$. The only difference between RC\_AB and RC\_Cm during the testing phase is the output matrix $W_o$. After training, we evaluate the performance of the RC by using different realizations and assess both one-step and autonomous predictions, with particular attention to the rogue wave occurrence.

\begin{figure}[tbhp]
    \centering
    \includegraphics[width=0.85\columnwidth]{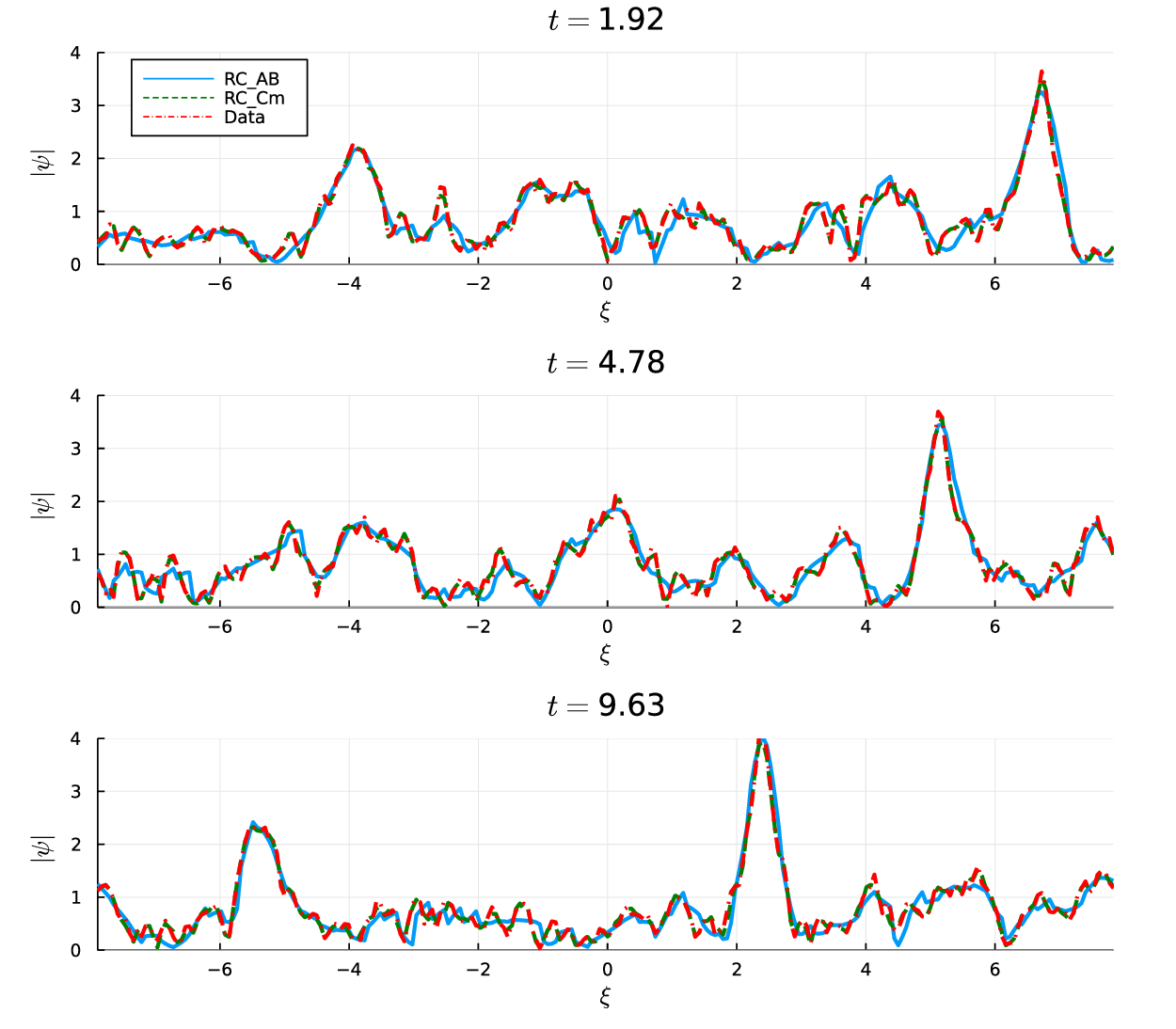}
    \caption[RC can capture the location of rogue waves in random ocean waves for one-step prediction.]{RC can capture the location of rogue waves in random ocean waves for one-step prediction. The plot shows the envelope profile at three distinct times $t=1.92,4.78,9.63$. }
    \label{fig:random_waves_onestep}
\end{figure}

Both RC models demonstrate reasonable agreement for one-step prediction, as illustrated in Fig.~\ref{fig:random_waves_onestep}, which shows one-step predictions at distinct time points. Note that time is presented in dimensionless units. All models successfully predict the occurrence and location of rogue waves, which is particularly noteworthy for RC\_AB, given that the shapes and locations of these rogue waves differ from those in the quasiperiodic training data. The ability of the parallel RC to accurately capture rogue waves at various locations can be attributed to the translation invariance enforced in its architecture. However, it is important to note that RC\_AB fails to capture the short-wave components of the wave envelope, likely due to the absence of such short waves in its training data.
\begin{figure}[tbhp]
    \centering
    \includegraphics[width=0.85\columnwidth]{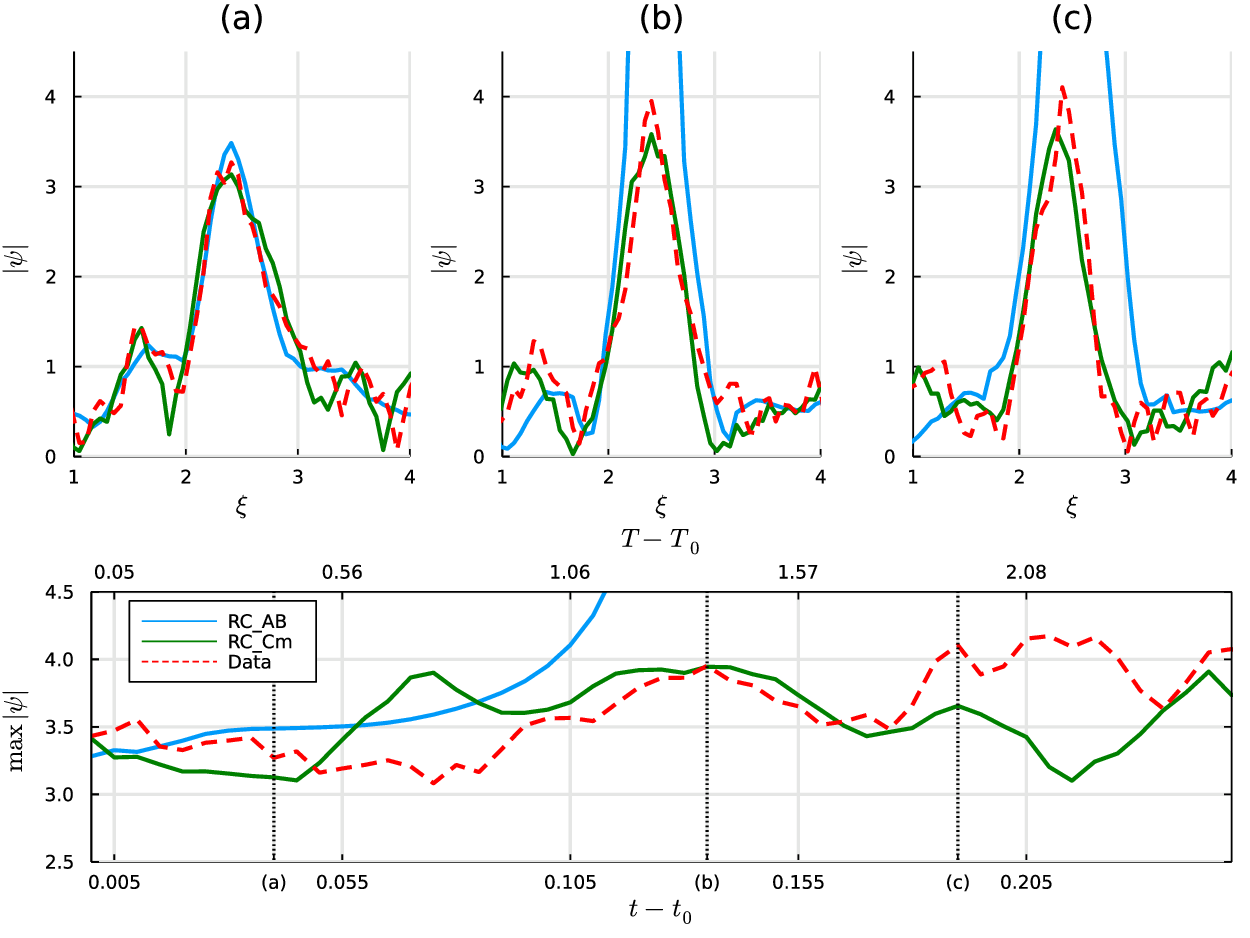}
    \caption[The autonomous prediction for random ocean wave.]{Autonomous prediction for random ocean waves. The bottom curve shows the maximal envelope amplitude with dimensionless time $t$ (bottom axis) and corresponding physical time $T$ (top axis). $t_0$ and $T_0$ are the start of forecasting time in dimensionless and physical time respectively. The upper figure corresponds to the profile of autonomous prediction at times corresponding to (a), (b), and (c). }
    \label{fig:auto_JS}
\end{figure}
For autonomous prediction, we initiate the forecasting process at $t_0$ which is 0.13 dimensionless time units prior to the rogue wave event, which corresponds to 1.32~s in physical time. The transformation between dimensionless time $t$ and physical time $T$ is given in Eq.~\eqref{eq:NLS_var_transform}. The results, presented in \ref{fig:auto_JS}, show that both RC\_Cm and RC\_AB exhibit reasonable agreement with the NLS simulation data in predicting the occurrence and location of rogue waves. Notably, RC\_Cm yields a more accurate estimate of the wave envelope's amplitude than RC\_AB.

We further evaluate the autonomous prediction capability of RC\_Cm using the task described in Subsection \ref{subsec:auto_pred}. The results, represented by the green curves in Figs.~\ref{fig:auto_profile} and \ref{fig:Hamiltonian}, indicate that RC\_Cm outperforms RC\_AB in autonomous prediction. This is evident from the close agreement between the solution profiles, particularly in terms of peak heights and locations, as shown in Fig.~\ref{fig:auto_profile}. Additionally, the potential energy of RC\_Cm aligns well with the NLS numerical simulation up to approximately 0.5 time units, while the Hamiltonian remains stable at around -0.5 until about one time unit.

These results highlight the versatility of the parallel RC architecture we proposed for random-wave simulations. Despite the absence of short-wave propagation in the training data for RC\_AB, the learned model still provides reasonable one-step and autonomous predictions. Furthermore, significant improvements could be achieved by incorporating random waves into the training data. It is worth noting that we attempted training using a dataset containing only random wave realizations; however, the resulting model did not perform well during the forecasting phase and is therefore not presented here. A possible explanation for this improvement is that RC\_AB is trained only on the phase space near homoclinic orbits, whereas RC\_Cm incorporates a broader region of phase space during training.

\section{Enhancing Autonomous Prediction \label{sec:auto_obs}}

In the previous sections, we observed that our RC model struggles to maintain long-term autonomous predictions. This degradation arises from two key issues: (1) the divergence of solutions due to error accumulation over time steps, and (2) the inability to preserve key invariants, such as the norm of the solution, which is crucial in Hamiltonian systems. To address these challenges and extend the prediction horizon, we propose two corrective approaches.

The first approach follows the method outlined in~\cite{Fan2020}, where measurements or real data points are intermittently introduced during autonomous prediction. Specifically, the input states in autonomous prediction are modified as
\begin{equation}
\tilde{u}^{(k)}_j = (1-\delta_j) u_j^{(k)} + \delta_j \hat{u}_j^{(k)},
\end{equation}
where $ u_j^{(k)} $ is the feedback input from the previous RC steps, $ \hat{u}_j^{(k)} $ is the measurement or the true state vector, and \( \delta_j \) is a Dirac step function that takes the value 1 at sparse, predetermined time steps and 0 otherwise. We adopt the simplest assimilation scheme in this work; for more variations of this method, we refer to~\cite{Fan2020}.

The second approach addresses the issue of norm loss in autonomous predictions, which can lead to unphysical solutions. To ensure that the global solution norm remains conserved throughout the autonomous prediction, we introduce a normalization step at each prediction step. This is achieved by rescaling the RC output as
\begin{equation}
\bar{y}_j^{(k)} = W_0 x_j^{(k)} \quad \text{and} \quad \hat{y}_j^{(k)} = c_j \bar{y}_j^{(k)},
\end{equation}
where the normalization factor $c_j$ is computed as
\begin{equation}
c_j = \frac{\|y_0\|}{\|\bar{y}_j\|},
\end{equation}
ensuring that the norm of the predicted solution remains consistent with the initial condition. Here, $\bar{y}_j$ is obtained by combining the individual components $\bar{y}_j^{(k)}$ from different reservoir units.

\begin{figure}[!tb]
    \centering
    \includegraphics[width=0.9\linewidth]{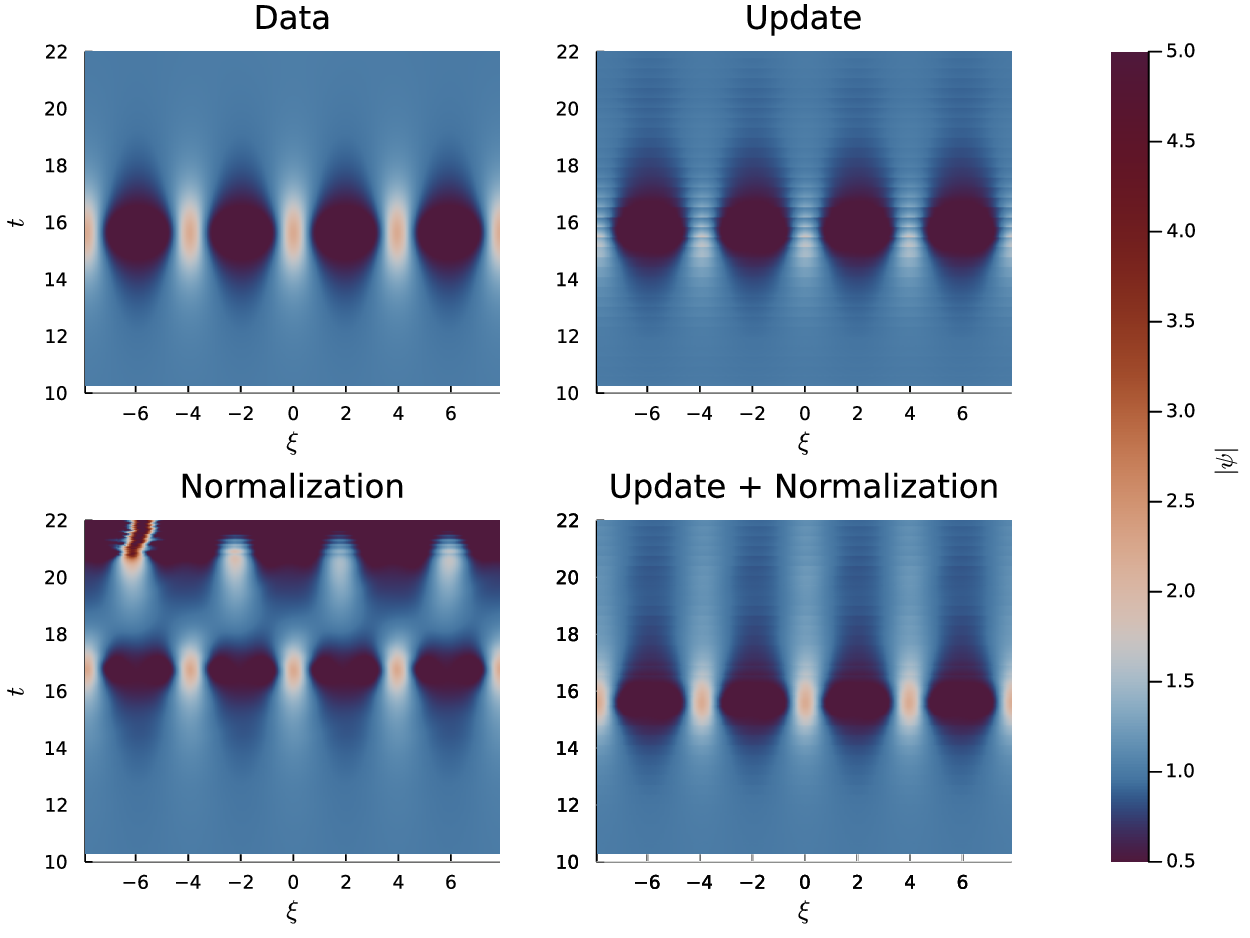}
    \caption[Heatmap of rogue wave dynamics.]{Heatmap of rogue wave dynamics. The upper left panel shows the ground-truth data. The remaining panels shows the autonomous RC\_AB prediction with modified procedures: partial update (upper right), normalization of the solution norm (lower left), and a combination of partial update and normalization of solution norm (lower right). The update and normalization procedures accurately capture both the amplitude and timing of the rogue wave.}
    \label{fig:heatmap_update}
\end{figure}

We investigate the case using RC\_AB, as described in Section~\ref{subsec:auto_pred}, by considering three scenarios: applying each stabilization approach separately and combining both methods. Compared to Section~\ref{subsec:auto_pred}, we initiate autonomous prediction at an earlier starting time, specifically \( t = 10 \), which is approximately 5.64 time units before the rogue wave formation. The update interval for the first method is set to 70 time steps, corresponding to one update every 0.35 time units.

Overall, all approaches successfully predict the occurrence of rogue waves. However, each method exhibits distinct limitations as clearly evident in Fig~\ref{fig:heatmap_update}. The update-only approach tends to underestimate the rogue wave amplitude, while the normalization approach predicts the rogue wave with a slight delay. In contrast, the combined approach accurately captures both the amplitude and timing of the rogue wave.

\begin{figure}[!tb]
    \centering
    \includegraphics[width=0.9\linewidth]{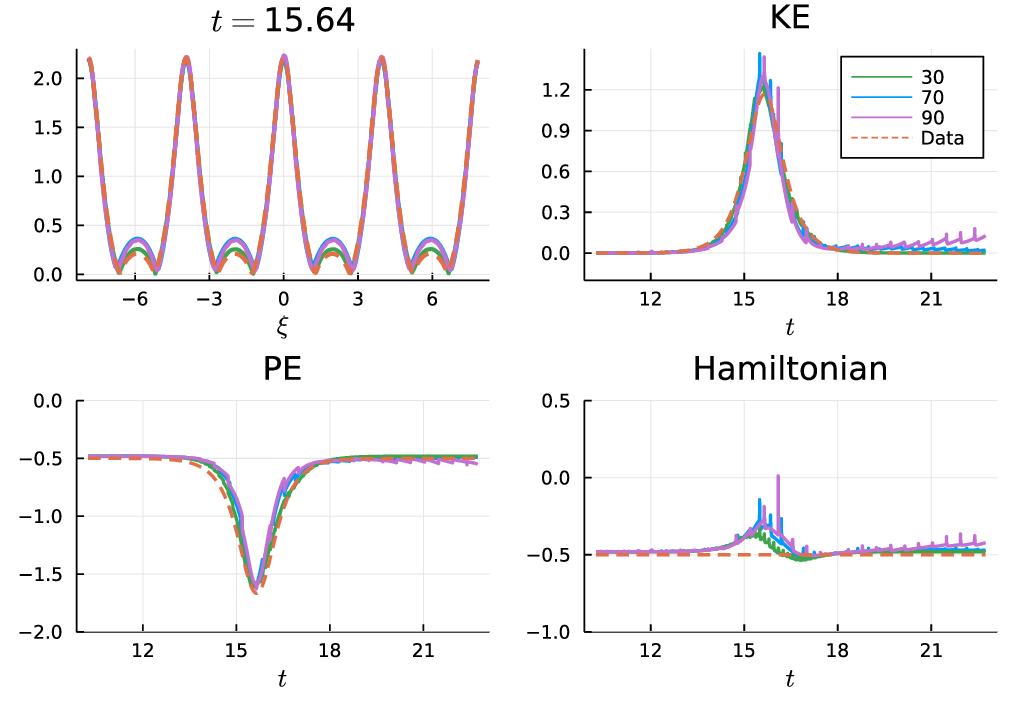}
    \caption[Sensitivity of the update interval: 30 (blue), 70 (green) and 90 (purple) time steps.]{Sensitivity of the update interval: 30 (blue), 70 (green), and 90 (purple) time steps. The red dashed line is the ground truth. The upper left panel displays the profile at the peak of the rogue wave. The remaining panels depict the system's energy components: Kinetic Energy (upper right), Potential Energy (lower left), and Total Energy/Hamiltonian (lower right).}
    \label{fig:hamil_update}
\end{figure}

Next, we examine the sensitivity of the update interval in the combined approach by considering three assimilation intervals: 30, 70, and 90 time steps. The top left panel of the Fig. \ref{fig:heatmap_update} shows the peak profile of the forecasted rogue wave, demonstrating close agreement in both shape and amplitude. The remaining panels display comparisons of potential and kinetic energy, which confirm that the combined approach effectively captures the energy exchange during rogue wave formation across all tested intervals.

Despite the accuracy in predicting the rogue wave profile, the length of the update interval significantly influences the reliability of the autonomous prediction over time. This is particularly evident when examining the Hamiltonian for the 90-time-step interval, which begins to deviate approximately 10 time units after the start of autonomous prediction. Longer update intervals tend to exhibit instability at earlier times. Overall, our combined approach significantly extends the prediction horizon of autonomous RC models.

\section{Discussion and Conclusion\label{chap4:sec:CONCLUSION}}

In this work, we have demonstrated that an RC model trained on a higher-order NLS breather can effectively forecast rogue wave simulations under various initial conditions, including random wave simulations. The prediction horizon varies depending on the initial autonomous prediction time \(t_0\), with predictions near the unstable manifold exhibiting a shorter horizon. This behavior contrasts with dissipative chaotic systems, where prediction statistics remain invariant with respect to the phase.

The RC model performs well in one-step predictions, which can be attributed to the loss function used during training (Eq.~\eqref{eq:Cost_Function}), which minimizes the error for one-step predictions. However, its predictive accuracy deteriorates during long-term autonomous forecasting, limiting its effectiveness as an independent time propagator. A key challenge we identified is that the autonomous RC dynamics lose the Hamiltonian property, even when trained on an integrable dataset. To address this issue, we have shown that incorporating partial measurement data and normalizing the solution norm can significantly improve the stability and accuracy of autonomous predictions.

We argue that the primary reason for RC's poor autonomous prediction performance lies in the presence of homoclinic orbits in the focusing NLS equation. Most of the training data reflect regular orbits, where rogue waves are obscured by the background. Rogue waves, which correspond to transversals through homoclinic orbits, are rare events that occur over longer time scales compared to the typical oscillatory motion of the NLS. As a result, the RC predominantly learns the "regular" dynamics of the NLS and struggles to capture the complex structure of the homoclinic orbit. Furthermore, in systems with multiple unstable modes, the unstable manifold has a complicated structure, making it challenging to provide training data that adequately span this manifold, which has a dimensionality corresponding to the number of unstable modes.

This observation may also explain why RC\_Cm outperforms RC\_AB. The training data for RC\_AB is limited to a small region close to the homoclinic orbit, meaning it primarily learns from data near this delicate structure. In contrast, RC\_Cm combines the RC\_AB training data with additional training from random wave simulations, which span a broader region of phase space, not confined to the vicinity of the homoclinic orbit. As a result, RC\_Cm is more robust during autonomous predictions, especially when deviations occur near the phase space surrounding the homoclinic orbit.

Regarding our initial motivation to evaluate RC's performance in predicting unseen data, we have observed a remarkable result: RC can synchronize well with a higher-order maximal intensity breather and random ocean waves, despite these not being part of the training data. Nevertheless, significant improvements, particularly in autonomous prediction, can be achieved by combining quasi-periodic waves with random waves in the training data. This finding suggests that to effectively learn the dynamics of spatio-temporal Hamiltonian systems, the training data must sufficiently sample the phase space, particularly the invariant manifold, a requirement that poses additional challenges for training.

An interesting follow-up question is how well a trained RC can capture the statistics of the NLS. Extreme events such as rogue waves are known to exhibit a heavy-tailed distribution due to modulation instability or nonlinearity \citep{Sapsis2021}. Given that RC has been shown to correctly capture climate statistics in several chaotic systems \citep{pathak_using_2017, Chattopadhyay2020, Doan2021}, it is natural to assess whether a trained RC can also accurately capture the statistics of the original dynamical system.

While our implementation employs a basic parallel RC architecture (Eqs.\ \eqref{eq:Internal_F}–\eqref{eq:Output_G}), recent RC variants offer promising directions for future refinement. Quantum RC \cite{fujii2017HarnessingDisorderedEnsemble}, Next Generation RC \cite{barbosa_learning_2022}, Stochastic RC \cite{ehlers2025StochasticReservoir}, and attention-based RC \cite{koster2025AttentionEnhancedReservoir} introduce novel approaches to reservoir dynamics and output layer design that may enhance chaotic system modeling. However, their comparative advantages for rogue wave forecasting remain to be systematically evaluated.

\section*{Data availability}

The data that support the findings of this study are available within the article or its supplementary materials.
    
	
\section*{Declaration of competing interest} 
	The authors declare that they have no known competing financial interests or personal relationships that could have appeared to influence the work reported in this paper.
	
	\section*{CRediT authorship contribution statement}
	The manuscript was written with contributions from all authors. All authors have given their approval to the final version of the manuscript.\\
	
	\textbf{ANH}: Methodology, Formal Analysis, Investigation, Software, Validation, Writing - Original Draft, Writing - Review \& Editing; {\textbf{HS}: Conceptualization, Validation, Supervision, Writing - Review \& Editing}.

\section*{Declaration of generative AI and AI-assisted technologies in the writing process}

During the preparation of this work the authors used Grammarly and ChatGPT in order to improve language and readability. After using these tools/services, the authors reviewed and edited the content as needed and took full responsibility for the content of the publication.

\section*{Acknowledgment}
We acknowledge the contribution of Khalifa University's high-performance computing and research computing facilities in providing computational resources used in this research. HS also acknowledges support from Khalifa University through a Faculty Start-Up Grant (No.\ 8474000351/FSU-2021-011), a Competitive Internal Research Awards Grant (No.\ 8474000413/CIRA-2021-065) and Research \& Innovation Grants (No.\ 8474000617/RIG-S-2023-031 and No.\ 8474000789/RIG-S-2024-070). 

\appendix
\section{Numerical Method for NLS Simulation \label{APP:Num_Method}} 

This appendix provides additional details on the numerical method used to generate simulation data. We adopt the Clean Numerical Simulation method \citep{hu_risks_2020} applied to the NLS equation (Eq.~\eqref{eq:NLS}).

To discretize the solution, we use a Taylor expansion of $\psi(\xi, t)$ in the variable $t$:
\begin{equation*}
    \psi(\xi, t + \Delta t) = \sum_{m=0}^{M} \psi^{[m]}(\xi, t)(\Delta t)^{m},
\end{equation*}
where $M$ is the Taylor order and $\Delta t$ is the time step. Additionally, we define:
\begin{equation*}
   \psi^{[m]}(\xi, t) = \frac{1}{m!} \frac{\partial^m \psi(\xi, t)}{\partial t^m}.
\end{equation*}
By substituting this ansatz into the NLS equation and matching the order of $(\Delta t)^m$, we obtain the recursion for the Taylor expansion:
\begin{equation}
   \psi^{[m+1]} = \frac{i}{m+1} \left( \frac{1}{2} \psi_{\xi\xi}^{[m]} + \sum_{j=0}^{m} \sum_{n=0}^{m-j} \bar{\psi}^{[j]} \psi^{[n]} \psi^{[m-j-n]} \right),
\end{equation}
where $\bar{\psi}$ denotes the complex e of $\psi$. The spatial derivatives are calculated in Fourier space. In our implementation, we choose $M = 6$.

\section{Maximum Local Lyapunov Exponent Calculation\label{APP:LLE}}

\begin{figure}
    \centering
    \includegraphics[width=0.8\linewidth]{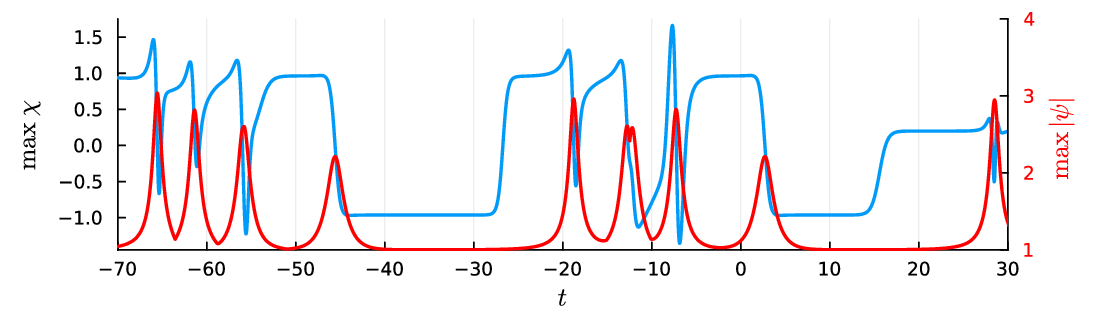}
    \caption{Temporal evolution of the maximum stretching number (local Lyapunov exponent, blue line) compared with the wave envelope amplitude (red line). Positive values indicate phases of instability growth, while negative values correspond to compression stages.}
    \label{fig:LLE}
\end{figure}

We computed the maximum stretching number \cite{Voglis_1994}, which is equivalent to the maximum local Lyapunov exponent \cite{eckhardt1993LocalLyapunova}, and defined as:
\begin{equation}
    \chi(t) = \frac{1}{\Delta t} \ln\frac{\left\Vert \delta \psi(\xi,t+\Delta t) \right\Vert}{\left\Vert \delta \psi(\xi,t) \right\Vert},
\end{equation}
where $\delta \psi$ denotes the deviation vector.

To evaluate this quantity, we solve the tangent (variational) equation associated with the nonlinear Schr\"odinger equation (NLS) \eqref{eq:NLS}:
\begin{equation}
    i \frac{\partial \delta \psi}{\partial t} + \frac{1}{2} \frac{\partial^2 \delta \psi}{\partial \xi^2} + 2|\psi|^2 \delta \psi + \psi^2 \overline{\delta \psi} = 0.
\end{equation}

The tangent dynamics were implemented using a Hamiltonian formulation with piecewise-constant coefficients, following \cite{skokos2010NumericalIntegration}. This method assumes the background field $\psi$ remains constant within each integration window of duration $\Delta t = 5 \times 10^{-3}$, which is significantly larger than the time step used for solving the NLS.

The computational procedure followed the approach in \cite{skokos2010LyapunovCharacteristic}, i.e.,
\begin{enumerate}
    \item Initialize the deviation vector $\delta \psi(\xi, 0)$ with uniformly distributed random values in $[-0.5, 0.5]$ and normalize such that $\left\Vert \delta \psi(\xi, 0) \right\Vert = 1$.
    \item At each step:
    \begin{enumerate}
        \item Integrate the variational equation over $\Delta t$.
        \item Compute the stretching number $\chi(t)$.
        \item Renormalize the deviation vector to unit norm: $\left\Vert \delta \psi(\xi,t) \right\Vert = 1$.
        \item Use the renormalized vector as the initial condition for the next interval.
    \end{enumerate}
\end{enumerate}

As we are interested solely in the maximal exponent, only a single deviation vector was evolved. Due to the exponential divergence along the most unstable direction, the initially random vector quickly aligns with this dominant mode.

For the training data shown in Fig.~\ref{fig:train_NLS}, the corresponding analysis in Fig.~\ref{fig:LLE} shows that the stretching number becomes positive during the onset of instability and negative during compression phases. The amplitude of the stretching number closely tracks the most active unstable mode and remains temporally synchronized with the envelope dynamics. Note that the largest Lyapunov exponent, defined as the long-time average of the stretching number, is not well-defined in this context due to persistent fluctuations over time.

\section{Numerical Simulation using JONSWAP Spectrum \label{App:JONSWAP}}

For ocean waves in deep water, the physical variables \(A\), \(X\), and \(T\) are related to the NLS variables by the following transformation (for a detailed derivation of the NLS for water waves, see \citep{kharifRogueWavesOcean2009}):
\begin{equation}
 t = \epsilon \omega_p T,  \quad \xi = 2 \epsilon K_p X, \quad \psi = \frac{K_p}{\epsilon \sqrt{2}} A^*, \label{eq:NLS_var_transform}
\end{equation}
where \(\epsilon\) comes from the scale-invariant NLS and will later be chosen to scale \(\psi\) such that \(\|\psi\| = 1\). The variables \(\omega_p\) and \(K_p\) represent the carrier wave period and the carrier wave number, respectively, and are related to the deep water dispersion relation: \(\omega_p = \sqrt{g K_p}\), where \(g\) is the gravitational constant. The variable \(A(X, T)\) denotes the wave envelope amplitude, where the first-order approximation is related to the surface elevation \(\eta\) by:
\begin{equation*}
    \eta(X, T) = A(X, T)\exp\left(i K_p X - i \omega_p T\right) + c.c.
\end{equation*}
In this test case, the surface elevation \(\eta\) is generated using the JONSWAP spectrum \citep{hasselmann1973MeasurementsWindwave}:
\begin{equation}
    S(f) = \frac{A_J}{f^5} \exp\left[-\frac{5}{4}\left(\frac{f}{f_p}\right)^{-4}\right] \gamma_S^r, \label{eq:JONSWAP}
\end{equation}
where \(f_p = \omega_p / (2\pi)\) and
\begin{align*}
        r &= \exp \left[-\frac{1}{2} \left(\frac{f / f_p - 1}{\sigma_S}\right)^2 \right], \\
    \sigma_S &= \begin{cases}
        0.07 & \text{if } f \leq f_p, \\
        0.09 & \text{if } f > f_p.
    \end{cases}
\end{align*}
The variable \(\gamma_S\) modifies the sharpness of the peak frequency. Its value is 3.3 for fetching limited sea in deep water and 1 for fully developed sea (Pierson-Moskowitz spectrum). The variable \(A_J\) scales the wave energy and is related to the significant wave height \(H_s\) and the peak period \(T_p\) as \citep{MACKAY2016}:
\begin{equation}
    A_J = \frac{H_s^2 f_p^4}{16 G_0(\gamma_S)}, \quad G_0(\gamma_S) = 0.1244 + 0.05 \gamma_S.
\end{equation}

The initial surface elevation is then generated as follows:
\begin{equation}
    \eta(X,0) = \sum_{i=1}^M C_i \cos(K_i X + \phi_i), \label{eq:init_condition_JONSWAP}
\end{equation}
where the amplitude is given by \(C_i = \sqrt{2 S(K_i) dK}\) and \(\phi_i\) are random phases uniformly distributed between \([0, 2\pi)\).

The procedure for generating the initial conditions is as follows. First, we specify the wave parameters \(H_s, T_p\). Then, using the NLS spatial variable \(\xi\), we define the spatial variable \(X\), the wavenumber space \(K\), and the frequency grid \(f\). The spectrum is calculated in the frequency space (Eq.\ \eqref{eq:JONSWAP}) and then transformed into the \(K\)-space while maintaining its total energy. Subsequently, the initial surface elevation is calculated using Eq.\ \eqref{eq:init_condition_JONSWAP}. The initial wave envelope is calculated as:
\[
 A(X, 0) = \frac{\eta(X, 0) + i \mathcal{H}[\eta(X, 0)]}{\exp(i K_p X)},
\]
where \(\mathcal{H}\) denotes the Hilbert transform. The next step is to transform the initial condition into the NLS variable, as described in Eq.\ \eqref{eq:NLS_var_transform}, and integrate the Cauchy problem in the NLS variable.

The simulations used the second-order split-step method, with the time step \(\Delta t = 10^{-3}\). It should be noted that the spatial NLS variable \(\xi\) used here corresponds identically to that described in Subsection \ref{Subsec:Train}. The parameters used for the JONSWAP spectrum are \(T_p = 8\) seconds, \(H_s = 8\) meters, and \(\gamma_S = 3.3\).

\bibliographystyle{elsarticle-num.bst}
\bibliography{Ref-RC}

\begin{thebibliography}{10}
\expandafter\ifx\csname url\endcsname\relax
  \def\url#1{\texttt{#1}}\fi
\expandafter\ifx\csname urlprefix\endcsname\relax\def\urlprefix{URL }\fi
\expandafter\ifx\csname href\endcsname\relax
  \def\href#1#2{#2} \def\path#1{#1}\fi

\bibitem{Liu200757}
P.~C. Liu, \href{https://hrcak.srce.hr/file/2720}{A chronology of freaque wave
  encounters}, Geofizika 24~(1) (2007) 57 -- 70.
\newline\urlprefix\url{https://hrcak.srce.hr/file/2720}

\bibitem{didenkulova_catalogue_2020}
E.~Didenkulova, Catalogue of rogue waves occurred in the {World} {Ocean} from
  2011 to 2018 reported by mass media sources, Ocean and Coastal Management 188
  (2020) 105076.
\newblock \href {http://dx.doi.org/10.1016/j.ocecoaman.2019.105076}
  {\path{doi:10.1016/j.ocecoaman.2019.105076}}.

\bibitem{kharif_physical_2003}
C.~Kharif, E.~Pelinovsky, Physical mechanisms of the rogue wave phenomenon,
  European Journal of Mechanics - B/Fluids 22~(6) (2003) 603--634.
\newblock \href {http://dx.doi.org/10.1016/j.euromechflu.2003.09.002}
  {\path{doi:10.1016/j.euromechflu.2003.09.002}}.

\bibitem{christou_field_2014}
M.~Christou, K.~Ewans, Field {Measurements} of {Rogue} {Water} {Waves}, Journal
  of Physical Oceanography 44~(9) (2014) 2317--2335.
\newblock \href {http://dx.doi.org/10.1175/JPO-D-13-0199.1}
  {\path{doi:10.1175/JPO-D-13-0199.1}}.

\bibitem{fedele_real_2016}
F.~Fedele, J.~Brennan, S.~Ponce~de Leon, J.~Dudley, F.~Dias, Real world ocean
  rogue waves explained without the modulational instability, Scientific
  Reports 6~(1) (2016) 27715.
\newblock \href {http://dx.doi.org/10.1038/srep27715}
  {\path{doi:10.1038/srep27715}}.

\bibitem{onorato_freak_2001}
M.~Onorato, A.~R. Osborne, M.~Serio, S.~Bertone, Freak {Waves} in {Random}
  {Oceanic} {Sea} {States}, Physical Review Letters 86~(25) (2001) 5831--5834.
\newblock \href {http://dx.doi.org/10.1103/PhysRevLett.86.5831}
  {\path{doi:10.1103/PhysRevLett.86.5831}}.

\bibitem{onorato_modulational_2006}
M.~Onorato, A.~R. Osborne, M.~Serio, Modulational {Instability} in {Crossing}
  {Sea} {States}: {A} {Possible} {Mechanism} for the {Formation} of {Freak}
  {Waves}, Physical Review Letters 96~(1) (2006) 014503.
\newblock \href {http://dx.doi.org/10.1103/PhysRevLett.96.014503}
  {\path{doi:10.1103/PhysRevLett.96.014503}}.

\bibitem{onorato_freak_2010}
M.~Onorato, D.~Proment, A.~Toffoli, Freak waves in crossing seas, The European
  Physical Journal Special Topics 185~(1) (2010) 45--55.
\newblock \href {http://dx.doi.org/10.1140/epjst/e2010-01237-8}
  {\path{doi:10.1140/epjst/e2010-01237-8}}.

\bibitem{Akhmediev2009}
N.~Akhmediev, J.~M. Soto-Crespo, A.~Ankiewicz, Extreme waves that appear from
  nowhere: {On} the nature of rogue waves, Physics Letters, Section A: General,
  Atomic and Solid State Physics 373~(25) (2009) 2137--2145.
\newblock \href {http://dx.doi.org/10.1016/j.physleta.2009.04.023}
  {\path{doi:10.1016/j.physleta.2009.04.023}}.

\bibitem{moslem2011SurfacePlasma}
W.~M. Moslem, P.~K. Shukla, B.~Eliasson, Surface plasma rogue waves,
  Europhysics Letters 96~(2) (2011) 25002.
\newblock \href {http://dx.doi.org/10.1209/0295-5075/96/25002}
  {\path{doi:10.1209/0295-5075/96/25002}}.

\bibitem{veldesElectromagneticRogueWaves2013}
G.~P. Veldes, J.~Borhanian, M.~McKerr, V.~Saxena, D.~J. Frantzeskakis,
  I.~Kourakis, Electromagnetic rogue waves in beam--plasma interactions,
  Journal of Optics 15~(6) (2013) 064003.
\newblock \href {http://dx.doi.org/10.1088/2040-8978/15/6/064003}
  {\path{doi:10.1088/2040-8978/15/6/064003}}.

\bibitem{yangAnalysisRogueWaves2021}
Y.~Yang, Y.-X. Gao, H.-W. Yang, Analysis of the rogue waves in the blood based
  on the high-order {{NLS}} equations with variable coefficients, Chinese
  Physics B 30~(11) (2021) 110202.
\newblock \href {http://dx.doi.org/10.1088/1674-1056/abff31}
  {\path{doi:10.1088/1674-1056/abff31}}.

\bibitem{li2024CharacteristicsCertain}
J.~Li, Z.-J. Yang, S.-M. Zhang, Characteristics of certain higher-order
  {{Hermite-cos-Gauss}} breathing solitons induced by the initial wavefront
  bending in optical media with nonlocal nonlinearity, Chaos, Solitons \&
  Fractals 187 (2024) 115338.
\newblock \href {http://dx.doi.org/10.1016/j.chaos.2024.115338}
  {\path{doi:10.1016/j.chaos.2024.115338}}.

\bibitem{liu2025RogueWaves}
Y.~Liu, Z.~Zhao, Rogue waves of the $(2+1)$-dimensional integrable reverse
  space--time nonlocal {{Schr{\"o}dinger}} equation, Theoretical and
  Mathematical Physics 222~(1) (2025) 34--52.
\newblock \href {http://dx.doi.org/10.1134/S0040577925010040}
  {\path{doi:10.1134/S0040577925010040}}.

\bibitem{slunyaevRogueWavesSea2023}
A.~V. Slunyaev, D.~E. Pelinovsky, E.~N. Pelinovsky, Rogue waves in the sea:
  Observations, physics, and mathematics, Physics-Uspekhi 66~(2) (2023)
  148--172.
\newblock \href {http://dx.doi.org/10.3367/UFNe.2021.08.039038}
  {\path{doi:10.3367/UFNe.2021.08.039038}}.

\bibitem{klein_deterministic_2020}
M.~Klein, M.~Dudek, G.~F. Clauss, S.~Ehlers, J.~Behrendt, N.~Hoffmann,
  M.~Onorato, On the {Deterministic} {Prediction} of {Water} {Waves}, Fluids
  5~(1) (2020) 9.
\newblock \href {http://dx.doi.org/10.3390/fluids5010009}
  {\path{doi:10.3390/fluids5010009}}.

\bibitem{Cousins2019}
W.~Cousins, M.~Onorato, A.~Chabchoub, T.~P. Sapsis, Predicting ocean rogue
  waves from point measurements: {An} experimental study for unidirectional
  waves, Physical Review E 99 (2019) 032201.
\newblock \href {http://dx.doi.org/10.1103/PhysRevE.99.032201}
  {\path{doi:10.1103/PhysRevE.99.032201}}.

\bibitem{Farazmand2017}
M.~Farazmand, T.~P. Sapsis, Reduced-order prediction of rogue waves in
  two-dimensional deep-water waves, Journal of Computational Physics 340 (2017)
  418--434.
\newblock \href {http://dx.doi.org/10.1016/j.jcp.2017.03.054}
  {\path{doi:10.1016/j.jcp.2017.03.054}}.

\bibitem{kagemoto_forecasting_2020}
H.~Kagemoto, Forecasting a water-surface wave train with artificial
  intelligence- {A} case study, Ocean Engineering 207 (2020) 107380.
\newblock \href {http://dx.doi.org/10.1016/j.oceaneng.2020.107380}
  {\path{doi:10.1016/j.oceaneng.2020.107380}}.

\bibitem{kagemoto_forecasting_2022}
H.~Kagemoto, Forecasting a water-surface wave train with artificial
  intelligence ({Part} 2) – {Can} the occurrence of freak waves be predicted
  with {AI}?, Ocean Engineering 252 (2022) 111205.
\newblock \href {http://dx.doi.org/10.1016/j.oceaneng.2022.111205}
  {\path{doi:10.1016/j.oceaneng.2022.111205}}.

\bibitem{breunung_data-driven_2023}
T.~Breunung, B.~Balachandran, Data-driven, high resolution ocean wave
  forecasting and extreme wave predictions, Ocean Engineering 268 (2023)
  113271.
\newblock \href {http://dx.doi.org/10.1016/j.oceaneng.2022.113271}
  {\path{doi:10.1016/j.oceaneng.2022.113271}}.

\bibitem{Jaeger2001}
H.~Jaeger, The "echo state" approach to analysing and training recurrent neural
  networks {\textendash} with an {Erratum} note, Tech. rep., German National
  Research Center for Information Technology, (2001).

\bibitem{Jaeger2004}
H.~Jaeger, H.~Haas, Harnessing {Nonlinearity}: {Predicting} {Chaotic} {Systems}
  and {Saving} {Energy} in {Wireless} {Communication}, Science 304~(5667)
  (2004) 78--80.
\newblock \href {http://dx.doi.org/10.1126/science.1091277}
  {\path{doi:10.1126/science.1091277}}.

\bibitem{Lukosevicius2009a}
M.~Luko{\v s}evi{\v c}ius, H.~Jaeger, Reservoir computing approaches to
  recurrent neural network training, Computer Science Review 3~(3) (2009)
  127--149.
\newblock \href {http://dx.doi.org/10.1016/j.cosrev.2009.03.005}
  {\path{doi:10.1016/j.cosrev.2009.03.005}}.

\bibitem{Chattopadhyay2020}
A.~Chattopadhyay, P.~Hassanzadeh, D.~Subramanian, Data-driven predictions of a
  multiscale {Lorenz} 96 chaotic system using machine-learning methods:
  {Reservoir} computing, artificial neural network, and long short-term memory
  network, Nonlinear Processes in Geophysics 27~(3) (2020) 373--389.
\newblock \href {http://dx.doi.org/10.5194/npg-27-373-2020}
  {\path{doi:10.5194/npg-27-373-2020}}.

\bibitem{pathak_using_2017}
J.~Pathak, Z.~Lu, B.~R. Hunt, M.~Girvan, E.~Ott, Using machine learning to
  replicate chaotic attractors and calculate {Lyapunov} exponents from data,
  Chaos: An Interdisciplinary Journal of Nonlinear Science 27~(12) (2017)
  121102, publisher: American Institute of Physics.
\newblock \href {http://dx.doi.org/10.1063/1.5010300}
  {\path{doi:10.1063/1.5010300}}.

\bibitem{Pathak2018}
J.~Pathak, B.~Hunt, M.~Girvan, Z.~Lu, E.~Ott, Model-{Free} {Prediction} of
  {Large} {Spatiotemporally} {Chaotic} {Systems} from {Data}: {A} {Reservoir}
  {Computing} {Approach}, Physical Review Letters 120~(2) (2018) 24102.
\newblock \href {http://dx.doi.org/10.1103/PhysRevLett.120.024102}
  {\path{doi:10.1103/PhysRevLett.120.024102}}.

\bibitem{Doan2021}
N.~A. Doan, W.~Polifke, L.~Magri, Short- {And} long-term predictions of chaotic
  flows and extreme events: {A} physics-constrained reservoir computing
  approach, Proceedings of the Royal Society A: Mathematical, Physical and
  Engineering Sciences 477 (2021) 2253.
\newblock \href {http://dx.doi.org/10.1098/rspa.2021.0135}
  {\path{doi:10.1098/rspa.2021.0135}}.

\bibitem{Rohm2021a}
A.~R{\"o}hm, D.~J. Gauthier, I.~Fischer, Model-free inference of unseen
  attractors: {Reconstructing} phase space features from a single noisy
  trajectory using reservoir computing, Chaos 31 (2021) 103127.
\newblock \href {http://dx.doi.org/10.1063/5.0065813}
  {\path{doi:10.1063/5.0065813}}.

\bibitem{gauthier_learning_2022}
D.~J. Gauthier, I.~Fischer, A.~R{\"o}hm, Learning unseen coexisting attractors,
  Chaos: An Interdisciplinary Journal of Nonlinear Science 32~(11) (2022)
  113107, publisher: American Institute of Physics.
\newblock \href {http://dx.doi.org/10.1063/5.0116784}
  {\path{doi:10.1063/5.0116784}}.

\bibitem{Huhn2021}
F.~Huhn, L.~Magri, Gradient-free optimization of chaotic acoustics with
  reservoir computing, Physical Review Fluids 7 (2022) 014402.
\newblock \href {http://dx.doi.org/10.1103/PhysRevFluids.7.014402}
  {\path{doi:10.1103/PhysRevFluids.7.014402}}.

\bibitem{Jiang2019}
J.~Jiang, Y.~C. Lai, Model-free prediction of spatiotemporal dynamical systems
  with recurrent neural networks: {Role} of network spectral radius, Physical
  Review Research 1 (2019) 033056.
\newblock \href {http://dx.doi.org/10.1103/PhysRevResearch.1.033056}
  {\path{doi:10.1103/PhysRevResearch.1.033056}}.

\bibitem{Pandey2020}
S.~Pandey, J.~Schumacher, Reservoir computing model of two-dimensional
  turbulent convection, Physical Review Fluids 5~(11) (2020) 113506.
\newblock \href {http://dx.doi.org/10.1103/PhysRevFluids.5.113506}
  {\path{doi:10.1103/PhysRevFluids.5.113506}}.

\bibitem{yao2022LearningOcean}
K.~Yao, E.~Forgoston, P.~Yecko, Learning ocean circulation models with
  reservoir computing, Physics of Fluids 34~(11) (2022) 116604.
\newblock \href {http://dx.doi.org/10.1063/5.0119061}
  {\path{doi:10.1063/5.0119061}}.

\bibitem{bonas2024CalibratedForecasts}
M.~Bonas, C.~K. Wikle, S.~Castruccio,
  \href{https://onlinelibrary.wiley.com/doi/abs/10.1002/env.2833}{Calibrated
  forecasts of quasi-periodic climate processes with deep echo state networks
  and penalized quantile regression}, Environmetrics 35~(1) (2024) e2833.
\newblock \href {http://dx.doi.org/10.1002/env.2833}
  {\path{doi:10.1002/env.2833}}.
\newline\urlprefix\url{https://onlinelibrary.wiley.com/doi/abs/10.1002/env.2833}

\bibitem{pathak_hybrid_2018}
J.~Pathak, A.~Wikner, R.~Fussell, S.~Chandra, B.~R. Hunt, M.~Girvan, E.~Ott,
  Hybrid forecasting of chaotic processes: {Using} machine learning in
  conjunction with a knowledge-based model, Chaos: An Interdisciplinary Journal
  of Nonlinear Science 28~(4) (2018) 041101, publisher: American Institute of
  Physics.
\newblock \href {http://dx.doi.org/10.1063/1.5028373}
  {\path{doi:10.1063/1.5028373}}.

\bibitem{Wikner2020}
A.~Wikner, J.~Pathak, B.~Hunt, M.~Girvan, T.~Arcomano, I.~Szunyogh,
  A.~Pomerance, E.~Ott, Combining machine learning with knowledge-based
  modeling for scalable forecasting and subgrid-scale closure of large,
  complex, spatiotemporal systems, Chaos 30~(5) (2020) 053111.
\newblock \href {http://dx.doi.org/10.1063/5.0005541}
  {\path{doi:10.1063/5.0005541}}.

\bibitem{tanakaRecentAdvancesPhysical2019}
G.~Tanaka, T.~Yamane, J.~B. H{\'e}roux, R.~Nakane, N.~Kanazawa, S.~Takeda,
  H.~Numata, D.~Nakano, A.~Hirose, Recent advances in physical reservoir
  computing: {{A}} review, Neural Networks 115 (2019) 100--123.
\newblock \href {http://arxiv.org/abs/1808.04962} {\path{arXiv:1808.04962}},
  \href {http://dx.doi.org/10.1016/j.neunet.2019.03.005}
  {\path{doi:10.1016/j.neunet.2019.03.005}}.

\bibitem{liang2024PhysicalReservoir}
X.~Liang, J.~Tang, Y.~Zhong, B.~Gao, H.~Qian, H.~Wu, Physical reservoir
  computing with emerging electronics, Nature Electronics 7~(3) (2024)
  193--206.
\newblock \href {http://dx.doi.org/10.1038/s41928-024-01133-z}
  {\path{doi:10.1038/s41928-024-01133-z}}.

\bibitem{everschor-sitte2024TopologicalMagnetic}
K.~{Everschor-Sitte}, A.~Majumdar, K.~Wolk, D.~Meier, Topological magnetic and
  ferroelectric systems for reservoir computing, Nature Reviews Physics 6~(7)
  (2024) 455--462.
\newblock \href {http://dx.doi.org/10.1038/s42254-024-00729-w}
  {\path{doi:10.1038/s42254-024-00729-w}}.

\bibitem{zhangLearningHamiltonianDynamics2021}
H.~Zhang, H.~Fan, L.~Wang, X.~Wang, Learning {{Hamiltonian}} dynamics with
  reservoir computing, Physical Review E 104~(2) (2021) 024205.
\newblock \href {http://dx.doi.org/10.1103/PhysRevE.104.024205}
  {\path{doi:10.1103/PhysRevE.104.024205}}.

\bibitem{pershin_training_2023}
A.~Pershin, C.~Beaume, K.~Li, S.~M. Tobias, Training a neural network to
  predict dynamics it has never seen, Physical Review E 107~(1) (2023) 014304.
\newblock \href {http://dx.doi.org/10.1103/PhysRevE.107.014304}
  {\path{doi:10.1103/PhysRevE.107.014304}}.

\bibitem{Kim2021}
J.~Z. Kim, Z.~Lu, E.~Nozari, G.~J. Pappas, D.~S. Bassett, Teaching recurrent
  neural networks to infer global temporal structure from local examples,
  Nature Machine Intelligence 3~(4) (2021) 316--323.
\newblock \href {http://dx.doi.org/10.1038/s42256-021-00321-2}
  {\path{doi:10.1038/s42256-021-00321-2}}.

\bibitem{manjunath_echo_2013}
G.~Manjunath, H.~Jaeger, Echo {State} {Property} {Linked} to an {Input}:
  {Exploring} a {Fundamental} {Characteristic} of {Recurrent} {Neural}
  {Networks}, Neural Computation 25~(3) (2013) 671--696.
\newblock \href {http://dx.doi.org/10.1162/NECO\_a\_0411}
  {\path{doi:10.1162/NECO\_a\_0411}}.

\bibitem{Barbosa2021}
W.~A. Barbosa, A.~Griffith, G.~E. Rowlands, L.~C. Govia, G.~J. Ribeill, M.~H.
  Nguyen, T.~A. Ohki, D.~J. Gauthier, Symmetry-aware reservoir computing,
  Physical Review E 104 (2021) 045307.
\newblock \href {http://dx.doi.org/10.1103/PhysRevE.104.045307}
  {\path{doi:10.1103/PhysRevE.104.045307}}.

\bibitem{Pyle2021}
R.~Pyle, N.~Jovanovic, D.~Subramanian, K.~V. Palem, A.~B. Patel, Domain-driven
  models yield better predictions at lower cost than reservoir computers in
  {Lorenz} systems, Philosophical Transactions of the Royal Society A:
  Mathematical, Physical and Engineering Sciences 379~(2194) (2021) 20200246.
\newblock \href {http://dx.doi.org/10.1098/rsta.2020.0246}
  {\path{doi:10.1098/rsta.2020.0246}}.

\bibitem{Akhmediev1987}
N.~N. Akhmediev, V.~M. Eleonskii, N.~E. Kulagin, Exact first-order solutions of
  the nonlinear {Schr{\"o}dinger} equation, Theoretical and Mathematical
  Physics 72~(2) (1987) 809--818.
\newblock \href {http://dx.doi.org/10.1007/BF01017105}
  {\path{doi:10.1007/BF01017105}}.

\bibitem{herbst_numerically_1989}
B.~M. Herbst, M.~J. Ablowitz, Numerically induced chaos in the nonlinear
  schr\"odinger equation, Physical Review Letters 62~(18) (1989) 2065--2068.
\newblock \href {http://dx.doi.org/10.1103/PhysRevLett.62.2065}
  {\path{doi:10.1103/PhysRevLett.62.2065}}.

\bibitem{belic_different_2022}
M.~R. Beli\'c, S.~N. Nikoli\'c, O.~A. Ashour, N.~B. Aleksi\'c, On different
  aspects of the optical rogue waves nature, Nonlinear Dynamics 108~(2) (2022)
  1655--1670.
\newblock \href {http://dx.doi.org/10.1007/s11071-022-07284-y}
  {\path{doi:10.1007/s11071-022-07284-y}}.

\bibitem{hu_risks_2020}
T.~Hu, S.~Liao, On the risks of using double precision in numerical simulations
  of spatio-temporal chaos, Journal of Computational Physics 418 (2020) 109629.
\newblock \href {http://dx.doi.org/10.1016/j.jcp.2020.109629}
  {\path{doi:10.1016/j.jcp.2020.109629}}.

\bibitem{Yuen1978}
H.~C. Yuen, W.~E. Ferguson, Relationship between {Benjamin}-{Feir} instability
  and recurrence in the nonlinear {Schr{\"o}dinger} equation, Physics of Fluids
  21~(8) (1978) 1275--1278.
\newblock \href {http://dx.doi.org/10.1063/1.862394}
  {\path{doi:10.1063/1.862394}}.

\bibitem{Erkintalo2011a}
M.~Erkintalo, K.~Hammani, B.~Kibler, C.~Finot, N.~Akhmediev, J.~M. Dudley,
  G.~Genty, Higher-order modulation instability in nonlinear fiber optics,
  Physical Review Letters 107~(25) (2011) 14--18.
\newblock \href {http://dx.doi.org/10.1103/PhysRevLett.107.253901}
  {\path{doi:10.1103/PhysRevLett.107.253901}}.

\bibitem{Chin2015}
S.~A. Chin, O.~A. Ashour, M.~R. Beli{\'c}, Anatomy of the {Akhmediev} breather:
  {Cascading} instability, first formation time, and {Fermi}-{Pasta}-{Ulam}
  recurrence, Physical Review E 92~(6) (2015) 063202.
\newblock \href {http://dx.doi.org/10.1103/PhysRevE.92.063202}
  {\path{doi:10.1103/PhysRevE.92.063202}}.

\bibitem{royModelfreePredictionMultistability2022a}
M.~Roy, S.~Mandal, C.~Hens, A.~Prasad, N.~V. Kuznetsov, M.~Dev~Shrimali,
  Model-free prediction of multistability using echo state network, Chaos: An
  Interdisciplinary Journal of Nonlinear Science 32~(10) (2022) 101104.
\newblock \href {http://dx.doi.org/10.1063/5.0119963}
  {\path{doi:10.1063/5.0119963}}.

\bibitem{Chin2016a}
S.~A. Chin, O.~A. Ashour, S.~N. Nikoli{\'c}, M.~R. Beli{\'c}, Maximal intensity
  higher-order {{Akhmediev}} breathers of the nonlinear {{Schr{\"o}dinger}}
  equation and their systematic generation, Physics Letters, Section A:
  General, Atomic and Solid State Physics 380~(43) (2016) 3625--3629.
\newblock \href {http://dx.doi.org/10.1016/j.physleta.2016.08.038}
  {\path{doi:10.1016/j.physleta.2016.08.038}}.

\bibitem{akhmediev_extremely_1991}
N.~Akhmediev, N.~Mitzkevich, Extremely high degree of {{N-soliton}} pulse
  compression in an optical fiber, IEEE Journal of Quantum Electronics 27~(3)
  (1991) 849--857.
\newblock \href {http://dx.doi.org/10.1109/3.81399}
  {\path{doi:10.1109/3.81399}}.

\bibitem{Kedziora2011}
D.~J. Kedziora, A.~Ankiewicz, N.~Akhmediev, Circular rogue wave clusters,
  Physical Review E - Statistical, Nonlinear, and Soft Matter Physics 84~(5)
  (2011) 056611.
\newblock \href {http://dx.doi.org/10.1103/PhysRevE.84.056611}
  {\path{doi:10.1103/PhysRevE.84.056611}}.

\bibitem{Fan2020}
H.~Fan, J.~Jiang, C.~Zhang, X.~Wang, Y.~C. Lai, Long-term prediction of chaotic
  systems with machine learning, Physical Review Research 2~(1) (2020) 012080.
\newblock \href {http://dx.doi.org/10.1103/PhysRevResearch.2.012080}
  {\path{doi:10.1103/PhysRevResearch.2.012080}}.

\bibitem{Sapsis2021}
T.~P. Sapsis, Statistics of {Extreme} {Events} in {Fluid} {Flows} and {Waves},
  Annual Review of Fluid Mechanics 53 (2021) 85--111.
\newblock \href {http://dx.doi.org/10.1146/annurev-fluid-030420-032810}
  {\path{doi:10.1146/annurev-fluid-030420-032810}}.

\bibitem{fujii2017HarnessingDisorderedEnsemble}
K.~Fujii, K.~Nakajima, Harnessing {{Disordered-Ensemble Quantum Dynamics}} for
  {{Machine Learning}}, Physical Review Applied 8~(2) (2017) 024030.
\newblock \href {http://dx.doi.org/10.1103/PhysRevApplied.8.024030}
  {\path{doi:10.1103/PhysRevApplied.8.024030}}.

\bibitem{barbosa_learning_2022}
W.~A.~S. Barbosa, D.~J. Gauthier, Learning spatiotemporal chaos using
  next-generation reservoir computing, Chaos: An Interdisciplinary Journal of
  Nonlinear Science 32~(9) (2022) 093137, publisher: American Institute of
  Physics.
\newblock \href {http://dx.doi.org/10.1063/5.0098707}
  {\path{doi:10.1063/5.0098707}}.

\bibitem{ehlers2025StochasticReservoir}
P.~J. Ehlers, H.~I. Nurdin, D.~Soh, Stochastic reservoir computers, Nature
  Communications 16~(1) (2025) 3070.
\newblock \href {http://dx.doi.org/10.1038/s41467-025-58349-6}
  {\path{doi:10.1038/s41467-025-58349-6}}.

\bibitem{koster2025AttentionEnhancedReservoir}
F.~K{\"o}ster, K.~Kanno, A.~Uchida, Attention-{{Enhanced Reservoir Computing}}
  as a {{Multiple Dynamical System Approximator}} (2025).
\newblock \href {http://arxiv.org/abs/2505.05852} {\path{arXiv:2505.05852}},
  \href {http://dx.doi.org/10.48550/arXiv.2505.05852}
  {\path{doi:10.48550/arXiv.2505.05852}}.

\bibitem{Voglis_1994}
N.~Voglis, G.~J. Contopoulos, Invariant spectra of orbits in dynamical systems,
  Journal of Physics A: Mathematical and General 27~(14) (1994) 4899.
\newblock \href {http://dx.doi.org/10.1088/0305-4470/27/14/017}
  {\path{doi:10.1088/0305-4470/27/14/017}}.

\bibitem{eckhardt1993LocalLyapunova}
B.~Eckhardt, D.~Yao, Local {{Lyapunov}} exponents in chaotic systems, Physica
  D: Nonlinear Phenomena 65~(1) (1993) 100--108.
\newblock \href {http://dx.doi.org/10.1016/0167-2789(93)90007-N}
  {\path{doi:10.1016/0167-2789(93)90007-N}}.

\bibitem{skokos2010NumericalIntegration}
C.~Skokos, E.~Gerlach, Numerical integration of variational equations, Physical
  Review E 82~(3) (2010) 036704.
\newblock \href {http://dx.doi.org/10.1103/PhysRevE.82.036704}
  {\path{doi:10.1103/PhysRevE.82.036704}}.

\bibitem{skokos2010LyapunovCharacteristic}
C.~Skokos, The {{Lyapunov Characteristic Exponents}} and {{Their Computation}},
  in: Dynamics of {{Small Solar System Bodies}} and {{Exoplanets}}, Springer,
  Berlin, Heidelberg, 2010, pp. 63--135.
\newblock \href {http://dx.doi.org/10.1007/978-3-642-04458-8\_2}
  {\path{doi:10.1007/978-3-642-04458-8\_2}}.

\bibitem{kharifRogueWavesOcean2009}
C.~Kharif, E.~Pelinovsky, A.~Slunyaev, Rogue {{Waves}} in the {{Ocean}},
  Advances in {{Geophysical}} and {{Environmental Mechanics}} and
  {{Mathematics}}, Springer, Berlin, Heidelberg, 2009.
\newblock \href {http://dx.doi.org/10.1007/978-3-540-88419-4}
  {\path{doi:10.1007/978-3-540-88419-4}}.

\bibitem{hasselmann1973MeasurementsWindwave}
K.~Hasselmann, T.~Barnett, E.~Bouws, H.~Carlson, D.~Cartwright, K.~Enke,
  J.~Ewing, H.~Gienapp, D.~Hasselmann, P.~Kruseman, A.~Meerburg, P.~Muller,
  D.~Olbers, K.~Richter, W.~Sell, H.~Walden, Measurements of wind-wave growth
  and swell decay during the joint north sea wave project ({{JONSWAP}}), Tech.
  rep., Deutches Hydrographisches Institut (1973).

\bibitem{MACKAY2016}
E.~Mackay, A unified model for unimodal and bimodal ocean wave spectra,
  International Journal of Marine Energy 15 (2016) 17--40.
\newblock \href {http://dx.doi.org/10.1016/j.ijome.2016.04.015}
  {\path{doi:10.1016/j.ijome.2016.04.015}}.

\end{thebibliography}

\end{document}